\ifdefined\TWOCOLUMN
\documentclass[conference]{IEEEtran}

\else
\documentclass[journal,onecolumn,twoside]{IEEEtran}
\fi

\usepackage{amssymb}
\usepackage{amsmath}
\usepackage{amsthm}
\usepackage{longfbox}
\usepackage{tikz}
\usetikzlibrary{fit,matrix}
\usetikzlibrary{patterns,snakes}
\usepackage{multirow}
\usepackage{makecell}

\usepackage{multicol}

\newtheorem{lemma}{Lemma}
\newtheorem{corollary}{Corollary}
\newtheorem{theorem}{Theorem}
\newtheorem{definition}{Definition}
\newtheorem{proposition}{Proposition}

\newtheorem{remark}{Remark}

\newtheorem{property}{Property}

\newcommand\given[1][]{\:#1\vert\:}

\DeclareMathAlphabet{\mathbsf}{OT1}{cmss}{bx}{n}
\DeclareMathAlphabet{\mathssf}{OT1}{cmss}{m}{sl}
\DeclareMathAlphabet{\mathcsf}{OT1}{cmss}{sbc}{n}

\newcommand{\svv}[1]{\mathbf{#1}}

\DeclareSymbolFont{bsfletters}{OT1}{cmss}{bx}{n}  
\DeclareSymbolFont{ssfletters}{OT1}{cmss}{m}{n}
\DeclareMathSymbol{\bsfGamma}{0}{bsfletters}{'000}
\DeclareMathSymbol{\ssfGamma}{0}{ssfletters}{'000}
\DeclareMathSymbol{\bsfDelta}{0}{bsfletters}{'001}
\DeclareMathSymbol{\ssfDelta}{0}{ssfletters}{'001}
\DeclareMathSymbol{\bsfTheta}{0}{bsfletters}{'002}
\DeclareMathSymbol{\ssfTheta}{0}{ssfletters}{'002}
\DeclareMathSymbol{\bsfLambda}{0}{bsfletters}{'003}
\DeclareMathSymbol{\ssfLambda}{0}{ssfletters}{'003}
\DeclareMathSymbol{\bsfXi}{0}{bsfletters}{'004}
\DeclareMathSymbol{\ssfXi}{0}{ssfletters}{'004}
\DeclareMathSymbol{\bsfPi}{0}{bsfletters}{'005}
\DeclareMathSymbol{\ssfPi}{0}{ssfletters}{'005}
\DeclareMathSymbol{\bsfSigma}{0}{bsfletters}{'006}
\DeclareMathSymbol{\ssfSigma}{0}{ssfletters}{'006}
\DeclareMathSymbol{\bsfUpsilon}{0}{bsfletters}{'007}
\DeclareMathSymbol{\ssfUpsilon}{0}{ssfletters}{'007}
\DeclareMathSymbol{\bsfPhi}{0}{bsfletters}{'010}
\DeclareMathSymbol{\ssfPhi}{0}{ssfletters}{'010}
\DeclareMathSymbol{\bsfPsi}{0}{bsfletters}{'011}
\DeclareMathSymbol{\ssfPsi}{0}{ssfletters}{'011}
\DeclareMathSymbol{\bsfOmega}{0}{bsfletters}{'012}
\DeclareMathSymbol{\ssfOmega}{0}{ssfletters}{'012}

\begin{document}

\tikzset{toprule/.style={%
        execute at end cell={%
            \draw [line cap=rect,#1] (\tikzmatrixname-\the\pgfmatrixcurrentrow-\the\pgfmatrixcurrentcolumn.north west) -- (\tikzmatrixname-\the\pgfmatrixcurrentrow-\the\pgfmatrixcurrentcolumn.north east);%
        }
    },
    bottomrule/.style={%
        execute at end cell={%
            \draw [line cap=rect,#1] (\tikzmatrixname-\the\pgfmatrixcurrentrow-\the\pgfmatrixcurrentcolumn.south west) -- (\tikzmatrixname-\the\pgfmatrixcurrentrow-\the\pgfmatrixcurrentcolumn.south east);%
        }
    }
}

\allowdisplaybreaks
%

\title{An Explicit Rate-Optimal Streaming
Code for Channels with Burst and Arbitrary Erasures}
\ifdefined\TWOCOLUMN
\author{
\IEEEauthorblockN{Elad Domanovitz}
\IEEEauthorblockA{
School of Electrical Engineering \\
Tel Aviv University\\
Tel Aviv-Yafo, Israel\\
                    E-mail: \texttt{domanovi@post.tau.ac.il}
}  \and
\IEEEauthorblockN{Silas L. Fong and Ashish Khisti}
\IEEEauthorblockA{
Department of Electrical and Computer Engineering\\
  University of Toronto\\
                    Toronto, ON M5S 3G4, Canada\\
                    E-mail: \texttt{\{silas.fong, akhisti\}@utoronto.ca}
}
}
\else
\author{Elad Domanovitz, Silas L. Fong and Ashish Khisti 
\thanks{E. Domanovitz and A. Khisti are with the Department of Electrical and Computer Engineering, University of Toronto, Toronto, ON M5S 3G4, Canada (E-mails: elad.domanovitz@utoronto.ca,akhisti@ece.utoronto.ca).}
\thanks{S. L. Fong is with Qualcomm Flarion Technologies, NJ 08807, USA
(E-mail: silas.fong@ieee.org).}
 \thanks{The material in this paper was presented in part at the 2019 IEEE Information Theory Workshop, Visby, Gotland, Sweden.}
}
\fi
\maketitle

\begin{abstract}
This paper considers the transmission of an infinite sequence of messages (a streaming source) over a packet erasure channel, where every source message must be recovered perfectly at the destination subject to a fixed decoding delay. While the capacity of a channel that introduces only bursts of erasures is well known, only recently, the capacity of a channel with either one burst of erasures or multiple arbitrary erasures in any fixed-sized sliding window has been established. However, the codes shown to achieve this capacity are either non-explicit constructions (proven to exist) or explicit constructions that require large field size that scales exponentially with the delay. This work describes an explicit rate-optimal construction for admissible channel and delay parameters over a field size that scales only quadratically with the delay.
\end{abstract}

\section{Introduction}

Real-time interactive video streaming became part of the day-to-day life of many people in the world. Unlike traditional traffic, which is not extremely sensitive to latency, real-time interactive video streaming is very sensitive to latency. According to \cite{cisco2018cisco}, while all IP video traffic will account for 82 percent of the overall traffic by 2022, the portion of real-time interactive video streaming is expected to grow dramatically in the upcoming years.

One of the fundamental requirements of a communication system is to handle interruptions that occur during the transmission of information. Such interruptions occur either due to the physical nature of the channel (for example, fading) or due to packet drops in an interim point of the link (for example, due to congestion or overload). In general, there are two main error control schemes in use: Automatic repeat request (ARQ) and forward error correction (FEC). 

When ARQ is used, the receiver acknowledges every received packet (or group of packets). If the transmitter does not receive an acknowledgment (after some predefined timeout), it sends this packet again. An inherent benefit of ARQ (compared to using FEC) is that the packets can be composed only of payload (no additional parity symbols are added). However, when considering low latency applications, it may not be an acceptable solution as its latency is at least three times the one-way delay of the link.

An alternative method for handling errors in the transmission is FEC. By ``sacrificing'' some throughput in advance, FEC has the potential to lower the recovery latency as it does not require signaling back to the transmitter. Traditionally, FEC is designed to maximize the error-correcting capabilities while ignoring the impact it may have on the latency. Two commonly used codes are Low-density parity-check (LDPC) \cite{gallager1962low,mackay1996near} and digital fountain codes \cite{luby2002lt,shokrollahi2006raptor}. The typical block length of these codes is very long (usually a few hundreds of symbols), resulting in high latency, which means they are not suitable for real-time interactive video streaming.

Martinian and Sunderberg first presented the challange of finding optimal low-latency codes in \cite{MartinianSundberg2004}. In their work, a new class of encoders was shown to have the shortest possible decoding delay required to correct all bursts of a given size with fixed redundancy. While assuming errors will happen in bursts is a realistic assumption for several real-life sources of errors (for example, an overflow of a buffer in any interim point of the transmission), it turns out that the error-correction capability of the burst-optimal codes deteriorates significantly even when a smaller amount of arbitrary erasures is introduced. Therefore, it is desirable that the code could handle both bursts and sporadic erasures efficiently.

Motivated by this goal, Badr, et al., \cite{BKTA2013} studied codes for a class of channels that introduce both burst and isolated erasures. They further presented the sliding-window burst erasure model and developed an upper bound for a case in which there are either a burst of $B$ erasures or a maximum of $N$ arbitrary erasures in any window of length $W$. However, an achievable scheme that achieves the upper bound was presented only for $R=1/2$.

Recently, two achievable schemes which achieve the upper bound defined in \cite{BKTA2013} (and thus show it is the capacity of this channel) were presented independently by Fong et al. in \cite{fong2019optimal} and Krishnan et al. in \cite{krishnan2018rate} for any rate. The proof in \cite{fong2019optimal} is existential (proving the existence of an appropriate generator matrix), while the field size requirements are large ($O{{T}\choose{N}}$). Recently, Dudzicz et al. in \cite{Dudzicz2019} showed an explicit (and systematic) construction for all rates greater than or equal $1/2$, albeit the required field size is larger than the non-explicit constructions of \cite{fong2019optimal}.

In \cite{krishnan2018rate}, an explicit construction based on linearized polynomials is presented. However, except for a small range of parameters, the field-size requirements are still large (scale exponentially with $T$). The field size was further addressed in \cite{KrishnanShuklaKumar2019}, in which a new rate-optimal code construction which covers all channel and delay parameters. The field size of this construction was showed to grow quadratically with the maximal delay constraint ($O(T^2)$). However, explicit constructions were presented only for specific cases. In this paper, we present an explicit construction that requires a field size which scales quadratically with the maximal delay constraint for any delay $T$, burst size $B$, or $N$ arbitrary erasures such the $T\geq B \geq N$.

The rest of this paper is organized as follows. Section~\ref{sec:StreamingCodes} provides basic definitions of streaming codes. Section~\ref{sec:channelModel} outlines the channel model used in this paper. Section~\ref{sec:Notation} provides basic definitions and properties of block codes that will be used to show that the suggested construction achieves capacity. Section~\ref{sec:codeConstruction} describes the process of designing the generator matrix of a block code that we later show is used to achieve the streaming capacity, with a field size that scales quadratically with the latency constraint. This construction is based on MDS codes; therefore, some of their properties are recalled in Section~\ref{sec:mdsCodes}. Section~\ref{sec:example} provides an example for capacity-achieving streaming code for a channel with $B=4,~N=3$ and a maximal decoding delay of $T=6$. Section~\ref{sec:proofOfThm1} provides a proof that the block code built using the procedure described in Section~\ref{sec:codeConstruction} can recover from any burst of $B$ erasures or $N$ sporadic erasures with maximal decoding delay of $T$. Finally, we note that \cite{KrishnanShuklaKumar2019} designed capacity-achieving streaming codes by specifying the parity-check matrix of these codes rather than describing its generator matrix. In section~\ref{sec:parityCheck}, we show that the parity-check matrix of the systematic version of the construction described in Section~\ref{sec:codeConstruction} meets the set of requirements defined in \cite{KrishnanShuklaKumar2019} for a parity-check matrix of a capacity-achieving streaming code.


\subsection{Notation}
\label{noation}
The set of non-negative integers is denoted by $\mathbb{Z}_+$. The set of $k$-dimensional row vectors over $\mathbb{F}$ is denoted by $\mathbb{F}^k$, and the set of $k\times n$ matrices over $\mathbb{F}$ is denoted by $\mathbb{F}^{k\times N}$. For any matrix $\svv{G}$, we let $\svv{G}^t$ and ${\rm rank}(\svv{G})$ denote respectively the transpose and the rank of G.

We denote by $\svv{G}_{\mathcal{A},\mathcal{B}}$ the sub-matrix generated from $\svv{G}$ by taking rows with indices in $\mathcal{A}$ and columns with indices in $\mathcal{B}$. We denote by $\svv{G}_{r_1:r_2,c_1:c_2}$ the ($r_2-r_1+1\times c_2-c_1+1$) sub-matrix generated from taking rows $r_1$ up to $r_2$ and columns $c_1$ up to $c_2$ from $\svv{G}$ where $r_1\geq 1$ and $c_1\geq 1$, i.e., we denote the index of the first row and column as ``1''. We denote by $\svv{G}_{:,c_1:c_2}$ the sub-matrix generated from taking columns $c_1$ up to $c_2$ from $\svv{G}$, and with $\svv{G}_{r_1:r_2,:}$ the sub-matrix formed from taking rows $r_1$ up to $r_2$ from $\svv{G}$. Further, denoting $\svv{G}^{-1}_{r_1:r_2,c_1:c_2}$ means that we take the sub-matrix defined above from  $\svv{G}^{-1}$.

\subsection{Streaming Codes}
Let ${\bf y}_i$ be the packet received by the destination at time $i$ for each $i\in\left\{0,1,\ldots,i+T\right\}$.
\label{sec:StreamingCodes}
\begin{definition}[\cite{badr2017layered}, Sec. II-B]
An $(n,k,T)_{\mathbb{F}}$-streaming code consists of the following:
\begin{enumerate}
    \item A sequence of source messages $\{{\bf s}_i\}_{i=0}^{\infty}$ where ${\bf s}_i \in \mathbb{F}^k$.
    \item An encoding function $f_i:\underbrace{\mathbb{F}^k\times\ldots\times\mathbb{F}^k}_{i+1~{\rm times}}\to\mathbb{F}^{n}$ for each $i\in\mathbb{Z}_{+}$, where $f_i$ is used by the source at time $i$ to encode ${\bf s}_i$ according to
    \begin{align*}
        {\bf x}_i=f_i\left({\bf s}_0,{\bf s}_1,\ldots,{\bf s}_i\right).
    \end{align*}
    \item A decoding function $\phi_{i+T}:\underbrace{\mathbb{F}^n\cup\{*\}\times\ldots\times\mathbb{F}^n\cup\{*\}}_{i+T+1~{\rm times}}\to\mathbb{F}^{n}$ for each $i\in\mathbb{Z}_{+}$ is used by the destination at time $i+T$ to estimate ${\bf s_i}$ according to
    \begin{align}
        \hat{\bf s}_i=\phi_{i+T}\left({\bf y}_0,{\bf y}_1,\ldots,{\bf y}_{i+T}\right).
    \end{align}
\end{enumerate}
\end{definition}

\begin{definition}
An $(n,k,m,T)_{\mathbb{F}}$-convolutional code is an $(n,k,T)_{\mathbb{F}}$-streaming code constructed as follows: 
\begin{enumerate}
    \item Let \mbox{$\svv{G}^{\rm conv}_0,\svv{G}^{\rm conv}_1,\ldots,\svv{G}^{\rm conv}_m$} be $m+1$ generator matrices in $\mathbb{F}^{k\times n}$.
    \item Then, for each $i\in\mathbb{Z}_{+}$
\begin{align}
    {\bf x}_i=\sum_{l=0}^{m}{\bf s}_{i-l}\svv{G}^{\rm conv}_l
\end{align}
where ${\bf s}_{-1}={\bf s}_{-2}=\ldots={\bf s}_{-m}={\bf 0}^{l \times k}$ by convention.
\end{enumerate} 
\end{definition}

\subsection{Channel Model}
\label{sec:channelModel}
The channel model considered in this work is the sliding-window burst erasure channel that was introduced by Badr et al.\ in \cite{BKTA2013}. This model introduces up to $B$ consecutive erasures, or $N$ arbitrarily positioned isolated erasures in any window of size $W$ among the sequence of transmitted packets ${\bf x}[t]$.

\begin{definition}
An erasure sequence is a binary sequence denoted by $e^{\infty}\triangleq\{e_i\}_{i=0}^{\infty}$ where \mbox{$e_i={\bf 1}\{{\rm erasure~occurs~at~time~i}\}$}.
\end{definition}

\begin{definition}
The mapping $g_{n}:\mathbb{F}^{n}\times \{0,1\}\to \mathbb{F}^{n}\cup\{*\}$ of an erasure channel is defined as
\begin{align}
    g_{n}({\bf x},e)=\begin{cases}
    {\bf x}~~~ {\rm if~}e=0,\\
    * ~~~ {\rm if~}e=1.
    \end{cases}
    \label{eq:g_n}
\end{align}
\end{definition}

\begin{definition}
A $(W,B,N)$-erasure sequence is an erasure sequence $e^{\infty}$ that satisfies the following: For each $i\in\mathbb{Z}_+$ and any window
\begin{align}
    \mathcal{W}_i\triangleq\left\{i,i+1,\ldots,i+W-1\right\},
\end{align}
either $N<\sum_{l\in\mathcal{W}_i}e_l\leq B$ holds with all the $1$'s in $\left(e_i,e_{i+1},\ldots,e_{i+W-1}\right)$ are located at consecutive positions or $\sum_{l\in\mathcal{W}_i}e_l\leq N$ hold with no restrictions on the locations of $1$'s.
\end{definition}

In other words, a $(W,B,N)$-erasure sequence introduces either one
burst erasure with length no longer than $B$ or multiple arbitrary
erasures with total count no larger than $N$ in any window
$\mathcal{W}_i$, $\forall i\in\mathbb{Z}_+$. The set of $(W, B, N)$-erasure
sequences is denoted by $\Omega^{\infty}(W,B,N)$.


We further assume that $W \geq T + 1$. In case where $B<W\leq T + 1$ we can achieve the capacity by reducing the effective delay to $T_{\rm eff} = W-1$ as discussed in \cite{badr2017layered}. Furthermore, the capacity is trivially zero if $W\leq B$ as an erasure sequence that erases all the channel packets becomes admissible.

Thus we can assume without loss of generality that
\begin{align}
    W>T\geq B\geq N \geq 1.
    \label{eq:WTBN}
\end{align}
For further details refer to Section I-B of \cite{fong2019optimal}.

The next definition defines a streaming code which is $(W, B, N)$-achievable.
\begin{definition}
An $(n, k, T)F$-streaming code is said to be $(W, B, N)$-achievable if the following holds for any $(W, B, N)$-erasure sequence $e^{\infty}\in\Omega^{\infty}(W,B,N)$: For all $i\in\mathbb{Z}_+$ and all ${\bf s}_i\in\mathbb{F}^k$, we have
\begin{align}
\hat{\bf s}_i={\bf s}_i
\end{align}
where
\begin{align}
   \hat{\bf s}_i & = \phi_{i+T}\left({\bf y}_0,{\bf y}_1,\ldots,{\bf y}_{i+T}\right) \nonumber \\
   &= \phi_{i+T}\left(g_n({\bf x}_0,e_0),\ldots,g_n({\bf x}_{i+T},e_{i+T}) \right)
\end{align}
\end{definition}

\begin{definition}
Fix any $(W, T, B, N)$ that satisfies \eqref{eq:WTBN}. The $(W, T, B, N)$-capacity, denoted by $C_{W,T ,B,N}$, is the supremum of the rates attained by $(n, k, T)_{\mathbb{F}}$-streaming codes that are $(W, B, N)$-achievable, i.e.,
\begin{align}
   C_{W,T ,B,N}\triangleq \sup \left\{\frac{k}{n}\given[\Big]{\rm There~exists ~an}~(W,B,N){\rm-achievable}~(n,k,T)_{\mathbb{F}}{\rm-streaming~code~for~some }~\mathbb{F} \right\}
\end{align}
\end{definition}

Recently, independent works in \cite{fong2019optimal}, \cite{krishnan2018rate} established that the capacity of $\mathcal{C}(W,B,N)$ is given by:
\begin{align}
     C_{W,T,B,N}=\frac{T-N+1}{T-N+B+1}.
\end{align}


\subsection{Block Codes}
\label{sec:Notation}
As we show next, linear block codes serve as a building block for capacity-achieving streaming codes. Therefore, we recall several definitions and properties of these codes that will be of use in the following Sections.

Let $y[i]$ be the symbol received by the destination at time $i$ for each $i\in\left\{0,1,\ldots,n-1\right\}$.

\begin{definition}
A point-to-point $(n,k)_{\mathbb{F}}$-block code consists of the following
\begin{enumerate}
    \item A sequence of $k$ symbols $\{u[l]\}_{l=0}^{k-1}$ where $u[l]\in\mathbb{F}$.
    \item A generator matrix $\svv{G}\in\mathbb{F}^{k\times n}$. The source codeword is generated according to
    \begin{align}
        \left[x[0]~x[1]~\ldots~x[n-1]\right]=\left[u[0]~u[1]~\ldots~u[k-1]\right]\svv{G}
    \end{align}
    \item A decoding function $\varphi_{l+n}:\mathbb{F}\cup\{*\}\times\ldots\mathbb{F}\cup\{*\}\to\mathbb{F}$ for each $l\in\{0,1,\ldots,k-1\}$, where $\varphi_{l+n}$ is used by the destination at time to estimate $u[l]$ according to
    \begin{align}
        \hat{u}[l]=\varphi_{l+n}(y[0],y[1],\ldots,y[n-1])
        \label{eq:16_1}
    \end{align}
\end{enumerate}
\end{definition}

When the code is systematic, the generator matrix associated with it can be expressed in the form of
\begin{align}
    \svv{G}=\left[\svv{I}_{k\times k} \given[\Big] \svv{P}_{k\times (n-k)}\right]
    \label{eq:Gsys}
\end{align}
\begin{definition} {\bf Punctured Code}

Let $\mathcal{C}$ be an $(n,k)_\mathbb{F}$ linear code. Given a subset $\mathcal{P}$ of $[0 : n-1]$, the code $\mathcal{C}$ punctured on the coordinates in $\mathcal{P}$, is the linear code of length  $(n - |\mathcal{P}|)$ obtained from $\mathcal{C}$ by deleting all the coordinates in $\mathcal{P}$. We denote the punctured code as $\mathcal{C}\given[\big]_{\mathcal{P}}$.
\end{definition}

\begin{definition}{\bf Shortened Code}

Let $\mathcal{C}$ be an $(n,k)_\mathbb{F}$ linear code. Given a subset $\mathcal{P}$ of $[0 : n-1]$, consider the subcode $\mathcal{C}^*$ achieved when assuming
\begin{align}
    u[i] = 0 ~~~\forall ~i \in \mathcal{P}.
\end{align}

Then by the phrase $\mathcal{C}$ shortened on the coordinates in $\mathcal{P}$, we will mean the linear code of length $(n - |\mathcal{P}|)$ obtained from $\mathcal{C^*}$ after puncturing on the coordinates given by $\mathcal{P}$. We denote the shortened code as $\mathcal{C}^{\mathcal{P}}$.
\end{definition}

\begin{definition}{\bf Parity-check matrix}
A parity-check matrix, $\svv{H}$ of a $(n,k)_\mathbb{F}$ linear code $\mathcal{C}$, is a generator matrix of the dual code, $\mathcal{C}^\perp$. This means that a codeword ${\bf c}$ is in $\mathcal{C}$ if and only if the matrix-vector product $\svv{H}{\bf c}^T=0$. When $\svv{G}$ is systematic the parity-check matrix can be expressed as
\begin{align}
    \svv{H}=\left[-\svv{P}_{k\times (n-k)}^T\given[\Big]\svv{I}_{(n-k)\times (n-k)} \right].
\end{align}
\end{definition}

We recall the following Theorem and Lemma with respect to the parity-check matrix.

\begin{theorem}[\cite{huffmanfundamentals}, Theorem 1.5.7]
Let $\mathcal{P}\in[0:n-1]$. Then
\begin{align}
    (\mathcal{C}{\given[\big]}_{\mathcal{P}})^{\perp}=(\mathcal{C}^{\perp})^{\mathcal{P}}
\end{align}
and
\begin{align}
    (\mathcal{C}^\mathcal{P})^{\perp}=(\mathcal{C}^{\perp})\given[\big]_{\mathcal{P}}.
\end{align}
    \label{thm:thmNik}
\end{theorem}
Alternatively, this Theorem states that puncturing the generator matrix of $\mathcal{C}$ is equivalent to shortening its parity-check matrix and that shortening the generator matrix of $\mathcal{C}$ is equivalent to puncturing its parity-check matrix.

In the sequel, we use the following Lemma to derive a set of conditions on the parity-check matrix.

\begin{lemma}[\cite{KrishnanShuklaKumar2019}, Lemma II.6]
Consider an $(n,k)_{\mathbb{F}}$ block code $\mathcal{C}$ having a parity-check
matrix $\svv{H}$. Let the subset $\mathcal{P}$ of $[0 : n-1]$ be
erased from $\mathcal{C}$ and let $i\in\mathcal{P}$. Then the $i$-th code symbol in a codeword can be recovered from the code symbols of the same codeword corresponding to coordinates in $\mathcal{P}^c$
if:
\begin{align*}
    {\bf h}_i \notin {\rm span} \left< \{ {\bf h}_j \}_{j\in\mathcal{P}\setminus\{i\}} \right>
\end{align*}
where ${\bf h}_j$ denoted the $j$-th column of $\svv{H}$.
    \label{lemma:lemNik}
\end{lemma}

In order to generate capacity-achieving streaming codes, a subclass of linear block codes is required, which are block codes that conform to a (stricter) decoding constraint.

\begin{definition}
A point-to-point $(n,k,T)_{\mathbb{F}}$-block code consists of the following
\begin{enumerate}
    \item A sequence of $k$ symbols $\{u[l]\}_{l=0}^{k-1}$ where $u[l]\in\mathbb{F}$.
    \item A generator matrix $\svv{G}\in\mathbb{F}^{k\times n}$. The source codeword is generated according to
    \begin{align}
        \left[x[0]~x[1]~\ldots~x[n-1]\right]=\left[u[0]~u[1]~\ldots~u[k-1]\right]\svv{G}
    \end{align}
    \item A decoding function $\varphi_{l+T}:\mathbb{F}\cup\{*\}\times\ldots\mathbb{F}\cup\{*\}\to\mathbb{F}$ for each $l\in\{0,1,\ldots,k-1\}$, where $\varphi_{l+T}$ is used by the destination at time $\min\{l+T,n-1\}$ to estimate $u[l]$ according to
    \begin{align}
        \hat{u}[l]=\begin{cases}
        \varphi_{l+T}(y[0],y[1],\ldots,y[l+T])~~~{\rm if}~l+T\leq n-1 \\
        \varphi_{l+T}(y[0],y[1],\ldots,y[n-1])~~~{\rm if}~l+T> n-1
        \end{cases}
        \label{eq:16}
    \end{align}
\end{enumerate}
\label{def:nkt_block}
\end{definition}

\begin{definition}
An $(n,k,T)_{\mathbb{F}}$-block code is said to be $(W,B,N)$-achievable if the following holds for any $(W,B,N)$-erasure sequence $e^{\infty}\in\Omega^{\infty}(W,B,N)$: Let
\begin{align}
    y[i]\triangleq g_1(x[i],e_i)
    \label{eq:17}
\end{align}
be the symbol received by the destination at time $i$ for each $i \in\left\{0,1,\ldots,n-1\right\}$ where $g_1$ is as defined in \eqref{eq:g_n}. For the $(n,k,T)_{\mathbb{F}}$-block code, we have
\begin{align}
\hat{u}[i] = u[i]
\end{align}
for all $i\in\left\{0,1,\ldots,k-1\right\}$ and all $u[i]\in\mathbb{F}$ where $\hat{u}$ is constructed according to \eqref{eq:16} and \eqref{eq:17}.
\end{definition}

\subsection{Generating streaming $(n,k,T)_{\mathbb{F}}$-streaming code which is $(W,B,N)$-achievable from $(n,k,T)_{\mathbb{F}}$-block code which is $(W,B,N)$-achievable}
\label{sec:standardDefOfBlockCode}

The following Lemma shows that given an $(n,k,T)_{\mathbb{F}}$-block code which is $(W,B,N)$-achievable, a corresponding $(n,k,n-1,T)_{\mathbb{F}}$-convolutional code (which is an $(n,k,T)_{\mathbb{F}}$-streaming code) which is $(W,B,N)$-achievable can be constructed.

\begin{lemma}[Lemma 1 in \cite{fong2019optimal}]
Given an $(n,k,T)_{\mathbb{F}}$-block code which is $(W, B, N)$-achievable, we can construct an $(n, k, n-1,T)_{\mathbb{F}}$ convolutional code which is $(W, B, N)$-achievable. More specifically, given that $\svv{G}$ is the generator matrix of the $(n,k,T)_{\mathbb{F}}$-block code where $g_{i,j}$ is the entry
situated in row $i$ and column $j$ of $\svv{G}$, we can construct the $n-1$ generator matrices of the $(n,k,n-1,T)_{\mathbb{F}}$-convolutional code as follows. For each $l\in\{0,1,\ldots,n-1\}$, construct
\begin{align}
    \svv{G}_l^{\rm conv}\triangleq
    \begin{cases}
        \left[\begin{tabular}{c c c}
         ${\bf 0}^{k\times l}$ & ${\rm diag}\left(g_{0,l},g_{1,l+1},\ldots,g_{k-1,l+k-1}\right)$ & ${\bf 0}^{k\times (n-k-l)}$
         \end{tabular}\right] ~~~~~ {\rm if}~0\leq l \leq n-k \\
         \\
         \left[
            \begin{tabular}{c c}
                \multirow{2}{*}{${\bf 0}^{k \times l}$} & ${\rm diag}\left(g_{0,l},g_{1,l+1},\ldots,g_{n-1,l+n-1}\right)$ \\
                 & ${\bf 0}^{k-n+l\times (n-l)}$
         \end{tabular}
         \right]~~~~~~~~~~~~~~~~~~~ {\rm if}~n-k\leq l \leq n-1.
    \end{cases}
\end{align}
We note that $\svv{G}=\sum_{l=1}^{n-1}\svv{G}_l^{\rm conv}$. In particular, if we let ${\bf s}_i\triangleq\left[s_i[0]~ s_i[1]~\cdots~s_i[k-1]\right]$ and let
\begin{align}
    \left[x_i[0]~x_{i+1}[1]~\cdots~x_{i+n-1}[n-1]\right]\triangleq\left[s_i[0]~s_{i+1}[1]~\cdots~s_{i+k-1}[k-1]\right]\svv{G},
\end{align}
i.e., we apply diagonal interleaving for all $i\in\mathbb{Z}_+$, then the symbols generated at time $i$ by the $(n, k, n-1,T)_{\mathbb{F}}$ convolutional code (which is an $(n,k,T)_{\mathbb{F}}$-streaming code) are
\begin{align}
    {\bf x}_i\triangleq\left[x_i[0]~x_i[1]~\cdots~x_i[n-1]\right].
\end{align}
\label{lem:lem2}
\end{lemma}

\subsection{Main result}
\label{sec:mainResult}
The following Theorem is proved in Section~\ref{sec:proofOfThm1}.
\begin{theorem}
\label{thm:thm1}
For any $(W,T,B,N)$ that meets~\eqref{eq:WTBN}, there exists an $(n,k,T)_{\mathbb{F}}$-block code with
\begin{align}
    R & =\frac{k}{n} \nonumber \\
    & = \frac{T-N+1}{T+B-N+1} \nonumber \\
    & = C_{W,T,B,N}
\end{align}
which is $(W,B,N)$-achievable where the field size scales quadratically with the delay constraint ($O(T^2)$).
\end{theorem}

Recalling Lemma~\ref{lem:lem2}, it follows that using an $(n,k,T)_{\mathbb{F}}$-block code which is $(W,B,N)$-achievable with field size that scales quadratically with the delay constraint ($O(T^2)$), an $(n,k,T)_{\mathbb{F}}$-streaming code which is $(W,B,N)$-achievable can be generated with the field size scales quadratically with the delay constraint ($O(T^2)$).

Thus, proving Theorem~\ref{thm:thm1} means that for any $(W,T,B,N)$ that meets~\eqref{eq:WTBN}, there exists an $(n,k,T)_{\mathbb{F}}$-streaming code which is $(W,B,N)$-achievable can be generated where the field size scales quadratically with the delay constraint ($O(T^2)$). Thus, for any $T,B,N$, which meet~\eqref{eq:WTBN}, the capacity $C_{W,T,B,N}$ can be achieved with field size that scales quadratically with the delay.

\section{Explicit Construction of capacity-achieving streaming code}
\label{sec:codeConstruction}



We start by recalling some standard definitions of MDS codes. We then present an explicit method of generating a generator matrix of a $(W,B,N)$-achievable block code with field size which scales quadratically with the delay constraint which is based on MDS codes and concludes that it can be used to generate a $(W,B,N)$-achievable streaming code with the same field size.

\subsection{MDS codes}
\label{sec:mdsCodes}
We show next that a building block in generating capacity-achieving streaming code is maximum distance separable (MDS) block code. Linear block codes that achieve equality in the Singleton bound are called MDS codes. A $k\times n$ matrix $\svv{G}$ over a finite field $\mathbb{F}_{q}$, with $k \leq n$, will be referred to as a generator matrix of an MDS matrix if any $k$ distinct columns of $\svv{G}$ form a linearly independent set.

We recall the following properties with respect to the generator matrix of MDS code.
\begin{property}[\cite{roth1989mds}]
$\svv{G}$ generates a (systemtic) $(n,k)_{\mathbb{F}}$ MDS code if and only if every square submatrix of $\svv{P}_{k\times (n-k)}$ (defined in \eqref{eq:Gsys}) is non-singular, i.e. each square submatrix is invertible and has full rank.
\label{def:Cauchy}
\end{property}
\begin{corollary}
We note that this Property means that any rectangular submatrix of
$\svv{P}_{k\times (n-k)}$ has also full rank which equals the
minimal dimension. \label{col:CauchyForRectangular}
\end{corollary}
Another Corollary we use is the following.
\begin{corollary}
Define $\mathcal{A}\subset\{1,\ldots,k\}$ and $\mathcal{B}\subset\{1,\ldots,B\}$. The sub matrix $\svv{P}_{\mathcal{A},\mathcal{B}}$ which is generated from taking rows which match the values in $\mathcal{A}$ and columns which match the values in $\mathcal{B}$ is non-singular. Thus, 
\begin{align}
    {\rm rank}(\svv{P}_{\mathcal{A},\mathcal{B}})=\min\left(|\mathcal{A}|,|\mathcal{B}|\right).
\end{align} 
\label{col:rankOFpartialP}
\end{corollary}
This corollary holds since performing column swap or row swap on the parity matrix of an MDS code results with an MDS code. We further note that
\begin{corollary}
From Lemma~\ref{lemma:lemNik} and Property~\ref{def:Cauchy} it follows that any group of $n-k$ vectors the parity matrix $\svv{H}$ of an $(n,k)$ MDS code is linearly independent. 
\label{col:anyNminKIsInd}
\end{corollary}

Next, we state the following corollaries which outline the outcome of puncturing and shortening an MDS code. 
\begin{corollary}
When $\mathcal{C}$ is an $(n,k)_{\mathbb{F}}$ MDS code, puncturing it over $\mathcal{P}$ results with $(n-|\mathcal{P}|,k)_{\mathbb{F}}$ MDS code.
\label{col:MDSpunc}
\end{corollary}

\begin{corollary}
When $\mathcal{C}$ is an $(n,k)_{\mathbb{F}}$ MDS code, shortening it over $\mathcal{P}$ results with an $(n-|\mathcal{P}|,k-|\mathcal{P}|)_{\mathbb{F}}$ MDS code.
\label{col:MDSshort}
\end{corollary}

We conclude with the following Proposition on the minimal size
required to generate an MDS code.

\begin{proposition}[\cite{macwilliams1977theory}]
The minimal field size required to for an $(n,k)_{\mathbb{F}}$ MDS code scales as $O(n)$.
\label{prop:CodeSize}
\end{proposition}
For example, $(n,k)$ Reed-Solomon code, which is an MDS code, requires a field of size $n$.


\subsection{Explicit construction of $(W,B,N)$-achievable $(n,k,T)_{\mathbb{F}}$-block code}
\label{sec:ExplConstuct}
Next we describe how to generate (an explicit) $\svv{G}$ which is the generator matrix of $(n,k,T)_{\mathbb{F}}$-block code with a field size which scales quadratically with the delay $T$. In Section~\ref{sec:proofOfThm1} we show that this code is $(W,B,N)$-achievable for any $(W,T,B,N)$ which meet \eqref{eq:WTBN}.

We define
\begin{align}
    k&=T-N+1 \nonumber \\
    n&=k+B
    \label{eq:def_k_n}
\end{align}
in the same manner as was defined in \cite{fong2019optimal}. The construction of the generator matrix is done in the following steps:
\begin{enumerate}
    \item We start with an $(n,k)_{\mathbb{F}_q}$ MDS code $\mathcal{C''}$ which has the following generator matrix\footnote{We note that $Y$ is not a constant element, but rather represent (potentially different) element taken from $\mathbb{F}_q$. See Section~\ref{sec:example} for a specific example.}
  \begin{align}
      \svv{G}''=
        \tikz[baseline=(M.west)]{%
    \node[matrix of math nodes,matrix anchor=west,left delimiter={[},right delimiter={]},ampersand replacement=\& ] (M) {
        1 \& 0 \& 0 \& 0 \& \cdots \& 0 \&  Y \& \cdots \& \cdots \& \cdots \& \cdots \& Y\\
        0 \& 1 \& 0 \& 0 \& \cdots \& 0 \& Y \& \cdots \& \cdots \& \cdots \& \cdots \& Y\\
        0 \& 0 \& \ddots \& 0 \& \cdots \& 0 \& Y \& \cdots \& \cdots \& \cdots \& \cdots \& Y \\
        \vdots \& \vdots \&  \& 1 \& \& \vdots \& \vdots \& \& \& \& \& \vdots\\
        \vdots \& \vdots \&  \& \& \ddots \& 0 \& \vdots \& \& \& \& \& \vdots \\
        0 \& 0 \& \cdots \& \cdots \& 0 \& 1 \& Y \& \cdots \& \cdots \& \cdots \& \cdots \& Y \\
        };
    \node[draw,color=blue,fit=(M-1-7)(M-6-12),inner sep=-1.3pt,label={$\svv{P}''_{k\times B}$}] {};
    }.
    \label{eq:matGDoublePrime}
\end{align}
First we note that following Property~\ref{prop:CodeSize}, the minimal field size for this code scales linearly with the delay $T$, i.e. $q=O(T)$. We further note that following Corollary~\ref{col:CauchyForRectangular}, each submatrix of matrix $\svv{P}''$ has a full rank.
\item We perform row operations to generate code $\mathcal{C'}$ with the generator matrix
\begin{equation}
    \begin{aligned}
    \svv{G'}= \tikz[baseline=(M.west)]{%
    \node[matrix of math nodes,matrix anchor=west,left delimiter={[},right delimiter={]},ampersand replacement=\& ] (M) {
        1 \& X \& \cdots \& X \& 0 \& 0 \& 0 \& \cdots \& 0 \& X \& \cdots \& X \\
        0 \& 1 \& X \& \cdots \& X \& 0 \& 0 \& \cdots \& 0 \& X \& \cdots \& \vdots \\
        0 \& 0 \& \ddots \& \ddots \& \ddots \& \ddots \& 0 \& \cdots \& 0 \&  X \& \cdots \&  X \\
        \vdots \& \vdots \&  \& 1 \&\ddots \&\ddots \& \ddots \& \& \vdots \& X \& \cdots \& X \\
        \vdots \& \vdots \&  \& \& \ddots \& X \& \cdots \& X \& 0 \& \vdots \& \& \vdots \\
        0 \& 0 \& \cdots \& \cdots \& 0 \& 1 \&  X \& \cdots \& X \& X \& \cdots \& X \\
        };
        \draw[decorate,decoration={brace,raise=4mm,amplitude=10pt,mirror}](M-6-1.west) -- node [below of=M,below=-3mm] {$k$} (M-6-6.east);
        \draw[decorate,decoration={brace,raise=4mm,amplitude=10pt,mirror}](M-6-7.west) -- node [below of=M,below=-3mm] {$N-1$} (M-6-9.east);
        \draw[decorate,decoration={brace,raise=4mm,amplitude=10pt,mirror}](M-6-10.west) -- node [below of=M,below=-3mm] {$B-N+1$} (M-6-12.east);
    }.
    \end{aligned}
    \label{eq:matGprime}
\end{equation}
where the goal is to ``spread'' $N-1$ parity symbols diagonally with the information symbols.
As it is easy to see that code $\mathcal{C'}$ (and also $\mathcal{C''}$) can recover from a burst of size $B$ starting at time $0$ only at time $T'=k+B-1$. It follows that if $B>N$ we have $T'>T$ therefore this code does not meet the required maximal delay constraint. Nevertheless, as we show next, it is an important interim step.

This ``spreading'' is achieved via successive row cancellation. Equivalently, it can be denoted as  $\svv{G'}=\svv{M}\svv{G''}$ where matrix $\svv{M}$ is an upper triangular matrix which is denoted as

\begin{equation}
    \svv{M}= \tikz[baseline=(M.west)]{%
    \node[matrix of math nodes,matrix anchor=west,left delimiter={[},right delimiter={]},ampersand replacement=\& ] (M) {
        1 \& Y' \& \cdots \& Y' \& 0 \& \cdots \& \cdots \& 0 \\
        0 \& 1 \& Y' \& \cdots \& Y' \& 0 \& \cdots \& 0 \\
        \vdots \& \cdots \& \ddots \& \ddots \& \ddots \& \ddots \& \ddots \& \vdots  \\
         0 \& \cdots \& 0 \& 1 \& Y' \& \cdots \& Y' \& 0 \\
        0 \& \cdots \& 0 \& 0 \& 1 \& Y' \& \cdots \& Y' \\
        0 \& \cdots \& 0 \& 0 \& 0 \& \ddots \& \ddots \& Y' \\
        0 \& \cdots \& 0 \& 0 \& 0 \& 0 \& 1 \& Y' \\
        0 \& \cdots \& 0 \& 0 \& 0 \& 0 \& 0 \& 1 \\
        };
        \draw[decorate,decoration={brace,raise=4mm,amplitude=10pt}](M-1-2.west) -- node [above of=M,above=-3mm] {$N-1$} (M-1-4.east);
    }.
    \label{eq:matM}
\end{equation}
where $Y'$ denotes a function of one of the $Y$ symbols.

Matrix $\svv{M}$ is an upper-triangular matrix thus it is a full-rank matrix and hence it is invertible. Since an erasure of any $l$ columns in $\svv{G'}$ can be translated to an erasure of $l$ columns in $\svv{G''}$ (by multiplying with the inverse of $\svv{M}$), the following property holds.
\begin{property}
\label{prop:prop1}
Block code $\mathcal{C'}$ with generator matrix $\svv{G'}$ is an $(n,k)_{\mathbb{F}_q}$ MDS code.
\end{property}
\item Finally, denoting with $\mathbb{F}_{q^2}$ the extension field of $\mathbb{F}_{q}$, we replace the $(B-N+1)\times (B-N+1)$ upper right matrix with $\alpha\cdot\svv{I}_{B-N+1}$ where $\alpha\in\mathbb{F}_{q^2}\setminus\mathbb{F}_q$  which generates the code $\mathcal{C}$ with the generator matrix
\begin{equation}
    \begin{aligned}
        \svv{G}=
        \tikz[baseline=(M.west)]{%
    \node[matrix of math nodes,matrix anchor=west,left delimiter={[},right delimiter={]},ampersand replacement=\& ] (M) {
        1 \& X \& \cdots \& X \& 0 \& 0 \& 0 \& \cdots \& 0 \& \alpha \& \cdots \& 0 \\
        0 \& 1 \& X \& \cdots \& X \& 0 \& 0 \& \cdots \& 0 \& 0 \& \ddots \& 0 \\
        0 \& 0 \& \ddots \& \ddots \& \ddots \& \ddots \& 0 \& \cdots \& 0 \&  0 \& \cdots \&  \alpha \\
        \vdots \& \vdots \&  \& 1 \&\ddots \&\ddots \& \ddots \& \& \vdots \& X \& \cdots \& X \\
        \vdots \& \vdots \&  \& \& \ddots \& X \& \cdots \& X \& 0 \& \vdots \& \& \vdots \\
        0 \& 0 \& \cdots \& \cdots \& 0 \& 1 \&  X \& \cdots \& X \& X \& \cdots \& X \\
        };
        \draw[decorate,decoration={brace,raise=4mm,amplitude=10pt,mirror}](M-6-1.west) -- node [below of=M,below=-3mm] {$k$} (M-6-6.east);
        \draw[decorate,decoration={brace,raise=4mm,amplitude=10pt,mirror}](M-6-7.west) -- node [below of=M,below=-3mm] {$N-1$} (M-6-9.east);
        \draw[decorate,decoration={brace,raise=4mm,amplitude=10pt,mirror}](M-6-10.west) -- node [below of=M,below=-3mm] {$B-N+1$} (M-6-12.east);
    }.
    \end{aligned}
    \label{eq:matG_1}
\end{equation}
We note that this operation increased the field size of the code, thus the minimal required field size now scales as $O(T^2)$.
\end{enumerate}


The generator matrix $\svv{G}$ is composed of the following three blocks
\begin{itemize}
    \item $\svv{G}_1$ - The upper left $k\times (k+N-1)$ sub-matrix of $\svv{G}$.
    \item $\svv{G}_2$ - The lower right $\left(k-(B-N+1)\right)\times \left(n-(B-N+1)\right)$ sub-matrix of $\svv{G}$.
    \item $\svv{G}_3$ - The upper right $(B-N+1)\times (B-N+1)$ sub-matrix of $\svv{G}$.
\end{itemize}
These blocks are depicted in \eqref{eq:matG} below.
\begin{equation}
    \begin{aligned}
    \svv{G}=
        \tikz[baseline=(M.west)]{%
    \node[matrix of math nodes,matrix anchor=west,left delimiter={[},right delimiter={]},ampersand replacement=\& ] (M) {
        1 \& X \& \cdots \& X \& 0 \& 0 \& 0 \& \cdots \& 0 \& \alpha \& \cdots \& 0 \\
        0 \& 1 \& X \& \cdots \& X \& 0 \& 0 \& \cdots \& 0 \& 0 \& \ddots \& 0 \\
        0 \& 0 \& \ddots \& \ddots \& \ddots \& \ddots \& 0 \& \cdots \& 0 \&  0 \& \cdots \&  \alpha \\
        \vdots \& \vdots \&  \& 1 \&\ddots \&\ddots \& \ddots \& \& \vdots \& X \& \cdots \& X \\
        \vdots \& \vdots \&  \& \& \ddots \& X \& \cdots \& X \& 0 \& \vdots \& \& \vdots \\
        0 \& 0 \& \cdots \& \cdots \& 0 \& 1 \&  X \& \cdots \& X \& X \& \cdots \& X \\
        };
    \node[draw,color=blue,fit=(M-1-1)(M-6-9),inner sep=-1.3pt,label={[xshift=-1cm, yshift=0cm]\textcolor{blue}{$\svv{G}_1$}}] {};
    \node[draw,color=red,densely dashed,fit=(M-4-4)(M-6-12),inner sep=1.5pt,label={[xshift=-0.5cm, yshift=-2.5cm]\textcolor{red}{$\svv{G}_2$}}] {};
    \node[draw,color=green,dashdotted,fit=(M-1-10)(M-3-12),inner sep=4pt,label={[xshift=0cm, yshift=0cm]\textcolor{green}{$\svv{G}_3$}}] {};
    }
    \end{aligned}.
    \label{eq:matG}
\end{equation}

Before showing that generator matrix $\svv{G}$ generates an $(n,k,T)_{\mathbb{F}_{q^2}}$-block code (where $q=O(T)$) which is $(W,B,N)$-achievable, we detnoe the following properties of $\svv{G}_1$ and $\svv{G}_2$.
\begin{property}
\label{prop:pro2}
Block $\svv{G}_1$ is a generator matrix of a $(k+N-1,k)_{\mathbb{F}_q}$ MDS code.
\end{property}
This can be viewed from Corollary~\ref{col:MDSpunc} and noting that $\svv{G}_1$ is generated by of puncturing $(B-N-1)$ columns from $G'$ which is a generator matrix of $(n,k)$ MDS code.
\begin{property}
\label{prop:pro3}
Block $\svv{G}_2$ is a generator matrix of an $(n-(B-N+1),k-(B-N+1))_{\mathbb{F}_q}$ MDS code.
\end{property}

This property holds since it can be assumed that $\svv{G}_2$ is generated from $\svv{G'}$ by assuming that the first $(B-N+1)$ symbols encodes using $\svv{G'}$ ($\{x_0,\ldots,x_{B-N}\}$) were received without errors (equivalently assume that $\{u_0,\ldots,u_{B-N}\}=\{0,\ldots,0\}$). Even though $\svv{G'}$ is not systematic, due to its structure receiving $\{x_0,\ldots,x_{B-N}\}$ without errors means that $\{u_0,\ldots,u_{B-N}\}$ can be decoded correctly and hence can be cancelled from the other received symbols. Therefore, the remaining code is an $(n-(B-N+1),k-(B-N+1))$ MDS code over $\mathbb{F}_q$.\footnote{This can be also viewed as shortening MDS code $\svv{C'}$ by the first $(B-N+1)$ information symbols.}



Following Properties~\ref{prop:pro2} and \ref{prop:pro3} we denote the codes induced by $\svv{G}_1$ and $\svv{G}_2$ as ${\rm MDS}_1$ and ${\rm MDS}_2$.
\begin{remark}
Using \eqref{eq:def_k_n} it follows that ${\rm MDS}_2$ is a $(T,T-B)_{\mathbb{F}_q}$ MDS code.
\label{rem:remark1}
\end{remark}
\begin{remark}
In case $T>B$, symbols $\{x[k],\ldots,x[T-B]\}$ are composed only of from $\{u[B-N+1],\ldots,u[k-1]\}$, i.e., these are the only symbols of ${\rm MDS}_2$ without contribution from $\{u[0],\ldots,u[B-N]\}$ (which are not information symbols of ${\rm MDS}_2$). These symbols are marked with a solid frame in \eqref{eq:rem2} below.
\label{rem:remark2}
\end{remark}
\begin{equation}
    \begin{aligned}
    \svv{G}=
        \tikz[baseline=(M.west)]{%
    \node[matrix of math nodes,matrix anchor=west,left delimiter={[},right delimiter={]},ampersand replacement=\& ] (M) {
        1 \& X \& \cdots \& X \& 0 \& 0 \& 0 \& \cdots \& 0 \& \alpha \& \cdots \& 0 \\
        0 \& 1 \& X \& \cdots \& X \& 0 \& 0 \& \cdots \& 0 \& 0 \& \ddots \& 0 \\
        0 \& 0 \& \ddots \& \ddots \& \ddots \& \ddots \& 0 \& \cdots \& 0 \&  0 \& \cdots \&  \alpha \\
        \vdots \& \vdots \&  \& 1 \&\ddots \&\ddots \& \ddots \& \& \vdots \& X \& \cdots \& X \\
        \vdots \& \vdots \&  \& \& \ddots \& X \& \cdots \& X \& 0 \& \vdots \& \& \vdots \\
        0 \& 0 \& \cdots \& \cdots \& 0 \& 1 \&  X \& \cdots \& X \& X \& \cdots \& X \\
        };
    \node[draw,color=red,densely dashed,fit=(M-4-4)(M-6-12),inner sep=1.5pt,label={[xshift=-0.5cm, yshift=-2.5cm]\textcolor{red}{$\svv{G}_2$}}] {};
    \node[draw,color=black,solid,fit=(M-1-7)(M-6-9),inner sep=0pt,label={[xshift=-0.5cm, yshift=-2.5cm]}] {};
    }
    \end{aligned}.
    \label{eq:rem2}
\end{equation}
\begin{remark}
Although $\svv{G}$ is not a systematic generator matrix, we note that $\tilde{\svv{G}}=\svv{M}^{-1}\svv{G}$ is a systematic generator matrix. Since $\svv{M}$ is an invertible (upper triangular) matrix, we prove in Section~\ref{sec:parityCheck} that the parity-check matrix of $\tilde{\svv{G}}$ meets a set of conditions defined in \cite{KrishnanShuklaKumar2019} on the parity-check matrix of a capacity-achieving code.
\label{rem:remark3}
\end{remark}

\subsection{Example}
\label{sec:example}
As an example, we look at the case where $B=4$, $N=3$ and $T=6$. The first phases of the generator matrix are generated over $GF(11)$. In the final stage we take an element from $GF(121)\setminus GF(11)$. The resulting generator matrix of the $(8,4,6)_{GF(121)}$ code is:
\begin{equation}
    \tikz[baseline=(M.west)]{%
    \node[matrix of math nodes,matrix anchor=west,left delimiter={(},right delimiter={)},ampersand replacement=\&,row 1/.style={bottomrule=thick}] (M) {
    T= \& 0 \& 1 \& 2 \& 3 \& 4 \& 5 \& 6 \& 7 \\
    \& 1 \& 10 \& 9 \& 0 \& 0 \& 0 \& \alpha \& 0 \\
    \& 0 \& 1 \& 9 \& 1 \& 0 \& 0 \& 0 \& \alpha \\
    \& 0 \& 0 \& 1 \& 6 \& 9 \& 0 \& 4 \& 8  \\
    \& 0 \& 0 \& 0 \& 1 \& 4 \& 1 \& 9 \& 8  \\
     };
    \node[draw,color=blue,fit=(M-2-2)(M-5-7),inner sep=-1.3pt] {};
    \node[draw,color=red,densely dashed,fit=(M-4-4)(M-5-9),inner sep=1.5pt] {};
    \node[draw,color=green,dashdotted,fit=(M-2-8)(M-3-9),inner sep=3pt] {};
    },
\label{eq:exampG}
\end{equation}
where
\begin{itemize}
    \item ${\rm MDS}_1$ (marked in blue and solid frame in \eqref{eq:exampG}) is a $(6,4)_{\rm GF(11)}$ MDS code.
    \item ${\rm MDS}_2$ (marked in red and dashed frame in \eqref{eq:exampG}) is a $(6,2)_{\rm GF(11)}$ MDS code.
\end{itemize}
We set (for example) $\alpha=0\cdot11+1\cdot11$ as the element from $GF(121)\setminus GF(11)$.\footnote{As one of the most common examples for field extension is generating the complex field from the real field, we borrow the complex field notation to indicate that $\alpha$ is from the extension field with no intersection with the base field (i.e., in the example of complex fields, it is equivalent to taking a pure imaginary number).} Next, we demonstrate the decoding process for several cases of erasures. We focus on decoding symbol $u[0]$ and show for that for any burst of $B=4$ symbols or $N=2$ random erasures, $u[0]$ can be recovered with maximal delay of $T=6$. 

As decoding symbol $u[0]$ when symbol $x[0]$ is not erased is trivial, we focus only on cases where $x[0]$ is erased:
\begin{itemize}
    \item A burst of size $B=4$ starting at time $0$
\begin{equation*}
        \tikz[baseline=(M.west)]{%
    \node[matrix of math nodes,matrix anchor=west,left delimiter={(},right delimiter={)},ampersand replacement=\&,row 1/.style={bottomrule=thick}] (M) {
    T= \& 0 \& 1 \& 2 \& 3 \& 4 \& 5 \& 6 \& 7 \\
    \& 1 \& 10 \& 9 \& 0 \& 0 \& 0 \& \alpha \& 0 \\
    \& 0 \& 1 \& 9 \& 1 \& 0 \& 0 \& 0 \& \alpha \\
    \& 0 \& 0 \& 1 \& 6 \& 9 \& 0 \& 4 \& 8  \\
    \& 0 \& 0 \& 0 \& 1 \& 4 \& 1 \& 9 \& 8  \\
     };
    \node[draw,color=red,densely dashed,fit=(M-4-4)(M-5-9),inner sep=1.5pt] {};
    \node[draw,color=green,dashdotted,fit=(M-2-8)(M-3-9),inner sep=3pt] {};
    \coordinate (x) at (M-2-2);\coordinate (y) at (M-5-2);\draw (x) -- (y);
    \coordinate (x) at (M-2-3);\coordinate (y) at (M-5-3);\draw (x) -- (y);
    \coordinate (x) at (M-2-4);\coordinate (y) at (M-5-4);\draw (x) -- (y);
    \coordinate (x) at (M-2-5);\coordinate (y) at (M-5-5);\draw (x) -- (y);
    },
\end{equation*}
    Using ${\rm MDS}_2$, $u[2]$ and $u[3]$ can be decoded at time $5$ since we have two linear independent equations from ${\rm MDS}_2$ which is $(6,2)_{\rm GF(11)}$ MDS code. Thus, $u[2]$ and $u[3]$ can be recovered at time $5$. Cancelling $u[2]$ and $u[3]$ from $x[6]$ results with decoding $u[0]$ at time $6$ as required.
    \item $N=3$ sporadic erasures where $x[6]$ is erased:
\begin{equation*}
        \tikz[baseline=(M.west)]{%
    \node[matrix of math nodes,matrix anchor=west,left delimiter={(},right delimiter={)},ampersand replacement=\&,row 1/.style={bottomrule=thick}] (M) {
    T= \& 0 \& 1 \& 2 \& 3 \& 4 \& 5 \& 6 \& 7 \\
    \& 1 \& 10 \& 9 \& 0 \& 0 \& 0 \& \alpha \& 0 \\
    \& 0 \& 1 \& 9 \& 1 \& 0 \& 0 \& 0 \& \alpha \\
    \& 0 \& 0 \& 1 \& 6 \& 9 \& 0 \& 4 \& 8  \\
    \& 0 \& 0 \& 0 \& 1 \& 4 \& 1 \& 9 \& 8  \\
     };
    \node[draw,color=blue,fit=(M-2-2)(M-5-7),inner sep=-1.3pt] {};
    \coordinate (x) at (M-2-2);\coordinate (y) at (M-5-2);\draw (x) -- (y);
    \coordinate (x) at (M-2-7);\coordinate (y) at (M-5-7);\draw (x) -- (y);
    \coordinate (x) at (M-2-8);\coordinate (y) at (M-5-8);\draw (x) -- (y);
    }.
\end{equation*}
    Using ${\rm MDS}_1$, which is $(6,4)_{\rm GF(11)}$ MDS code, all information symbols can be decoded at time $4$ thus meeting the delay constraint $T$.
    \item $N=3$ sporadic erasures where $x[6]$ is not erased:
\begin{equation}
        \tikz[baseline=(M.west)]{%
    \node[matrix of math nodes,matrix anchor=west,left delimiter={(},right delimiter={)},ampersand replacement=\&,row 1/.style={bottomrule=thick}] (M) {
    T= \& 0 \& 1 \& 2 \& 3 \& 4 \& 5 \& 6 \& 7 \\
    \& 1 \& 10 \& 9 \& 0 \& 0 \& 0 \& \alpha \& 0 \\
    \& 0 \& 1 \& 9 \& 1 \& 0 \& 0 \& 0 \& \alpha \\
    \& 0 \& 0 \& 1 \& 6 \& 9 \& 0 \& 4 \& 8  \\
    \& 0 \& 0 \& 0 \& 1 \& 4 \& 1 \& 9 \& 8  \\
     };
    \node[draw,color=blue,fit=(M-2-2)(M-5-7),inner sep=0pt] {};
    \node[draw,color=blue,densely dashed,fit=(M-3-3)(M-5-7),inner sep=-2pt] {};
    \node[draw,color=magenta,densely dashed,fit=(M-3-8)(M-5-8),inner sep=-2pt] {};
    \coordinate (x) at (M-2-2);\coordinate (y) at (M-5-2);\draw (x) -- (y);
    \coordinate (x) at (M-2-6);\coordinate (y) at (M-5-6);\draw (x) -- (y);
    \coordinate (x) at (M-2-7);\coordinate (y) at (M-5-7);\draw (x) -- (y);
    }.
  \label{eq:exmp_N3_x6not}
  \end{equation}
    We note that the $(3\times 5)$ lower right matrix of $\svv{G}_1$ (marked in dashed blue in \eqref{eq:exmp_N3_x6not}) is a generator matrix of $(5,3)_{\rm GF(11)}$ MDS code which is the outcome of shortening $\svv{G}_1$ over $\mathcal{P}=\{0\}$ (denoted as ${\rm MDS}_1^1$). Receiving symbols $\{x[1],x[2],x[3]\}$ with no erasures means that bringing ${\rm MDS}_1^1$ to an equivalent row echelon form results with three linear combinations of $u[j]\in\{u[1],u[2],u[3]\}$ and $u[0]$ (with coefficients taken from ${\rm GF(11)}$) where each combination consists only of $u[j]$ and $u[0]$. Denoting these combination as  $\{\tilde{u}[1],\tilde{u}[2],\tilde{u}[3]\}$, using $\tilde{u}[2]$ and $\tilde{u}[3]$, $u[2]$ and $u[3]$ can be cancelled from $x[6]$. Since we assume that the element $\alpha=0\cdot11+1\cdot11\in\mathbb{F}_{{\rm GF(121)}\setminus\mathbb{\rm GF(11)}}$, it is guaranteed that it is not nulled out (since the column operations are performed over ${\rm GF(11)}$). Hence, $u[0]$ can be decoded at time $6$ as required.

    Equivalently, the dashed part of ${\bf g}_6$ (the part of $x[6]$ which is a function of $\{u[1],u[2],u[3]\}$) is in the span of ${\rm MDS}_1^1$. Since we have three symbols from ${\rm MDS}_1^1$, the dashed part of ${\bf g}_6$ can be cancelled. Since we assume that the element $\alpha=0\cdot11+1\cdot11\in\mathbb{F}_{{\rm GF(121)}\setminus\mathbb{\rm GF(11)}}$, it is guaranteed that it is not nulled out (since column operations are performed over ${\rm GF(11)}$). Hence, $u[0]$ can be recovered.
\end{itemize}
The decoding of $u[i]\in\{u[1],u[2],u[3]\}$ is done in a similar manner where we assume by induction that $\{u[0],\ldots,u[i-1]\}$ have already been recovered by time $T+i$. The following $(8,4,6)_{GF(121)}$-streaming code is then generated from this $(8,4,6)_{GF(121)}$ block code using diagonal interleaving. An example of a single diagonal is given in Table~\ref{tale:846Example} below.
\begin{table}[ht]
\begin{center}
\begin{tabular}{|c|c|c|c|c|c|c|c|}
    \hline
    ${\bf x}_i$ & ${\bf x}_{i+1}$ & ${\bf x}_{i+2}$ & ${\bf x}_{i+3}$ & ${\bf x}_{i+4}$ & ${\bf x}_{i+5}$ & ${\bf x}_{i+6}$ & ${\bf x}_{i+7}$ \\ \hline
    $s_i[0]$ & & & & & & &  \\ \hline
    & $10s_i[0]+s_{i+1}[1]$ & & & & & & \\ \hline
    & & \makecell{$9s_i[0]+9s_{i+1}[1]$\\+$s_{i+2}[2]$} & & & & & \\ \hline
    & & & \makecell{$s_{i+1}[1]+6s_{i+2}[2]$\\$+s[i+3]$} & & & & \\ \hline
    & & & & $9s_{i+2}[2]+4s[i+3]$ & & &\\ \hline
    & & & & & $s_{i+3}[3]$ & & \\ \hline
    & & & & & & \makecell{$11s_{i}[0]+4s_{i+2}[2]$\\$+9s_{i+3}[3]$} & \\ \hline
    & & & & & & & \makecell{$11s_{i+1}[1]+8s_{i+2}[2]$\\$+8s_{i+3}[3]$} \\ \hline
\end{tabular}.
\end{center}
\caption{An example of applying diagonal interleaving to generate $(8,4,6)_{GF(121)}$-streaming code for a $(8,4,6)_{GF(121)}$ block code.}
\label{tale:846Example}
\end{table}

\section{Proof Of Theorem~\ref{thm:thm1}}
\label{sec:proofOfThm1}
We prove next that the $(n,k,T)_{\mathbb{F}_{q^2}}$ block code described above is $(W,B,N)$ achievable. The field size of this construction scales as $O(T^2)$. We therefore describe the decoding function $\varphi_{l+T}$ defined in Definition~\ref{def:nkt_block} and prove it can recover any $u[l]$ (for any $l\in\{0,\ldots,k-1\}$) with maximal delay of $T$.

\begin{itemize}
    \item Decoding information symbols $\{u[0],\ldots,u[B-N]\}$\newline 
    We analyze the two different types of erasures:
    \begin{itemize}
        \item A Burst of length B starting at time $i$\newline
        {\bf Decoding information symbol $u[0]$} \newline
        Following Property~\ref{prop:pro3} and Remark~\ref{rem:remark1} we recall that ${\rm MDS}_2$ is a \mbox{$(T,T-B)_{\mathbb{F}_q}$} MDS code. A burst of $B$ symbols starting at time $0$ means that symbols $\{x[k],\ldots,x[T+B]\}$ are not erased. Recalling Remark~\ref{rem:remark2}, these symbols don't have an interference from information symbols $\{u[0],\ldots,u[B-N]\}$. Therefore, information symbols $\{u[B-N+1],\ldots,u[k]\}$ can be recovered using ${\rm MDS}_2$.

        Noting that $x[T]$ is composed of $u[0]$ and $\{u[B-N+1],\ldots,u[k]\}$, after symbols $\{u[B-N+1],\ldots,u[k]\}$ are decoded, they can cancelled from $x[T]$ thus $u[0]$ can be decoded.\footnote{In case $T=B$ it can be shown that $\svv{G}_3=\alpha\cdot\svv{I}_{B-N+1}$ hence $s[0]$ can be recovered directly from $x[T]$.}

        {\bf Decoding information symbol $u[i]\in\{u[1],\ldots,u[B-N]\}$}\newline
        Noting that $x[i+T]$ is composed of $u[i]$ and $\{u[B-N+1],\ldots,u[k]\}$, we again perform the decoding in two steps:
        \begin{itemize}
        \item Decoding of $\{u[B-N+1],\ldots,u[k]\}$ using ${\rm MDS}_2$.
        \item Decoding of $u[i]$ from $x[i+T]$ by canceling  $\{u[B-N+1],\ldots,u[k]\}$ from it.
        \end{itemize}
        We first assume by induction that symbols $\{u[0],\ldots, u[i-1]\}$ have already been recovered by time $T+i$, thus we cancel information symbols $\{u[0],\ldots, u[i-1]\}$ from $\{x[i+B],\ldots,x[i+T-1]\}$. Denoting $\{\tilde{x}[i+B],\ldots,\tilde{x}[i+T-1]\}$ as the symbols after the cancellation, we note that $\{\tilde{x}[i+B],\ldots,\tilde{x}[i+T-1]\}$ belong to ${\rm MDS}_2$.

        Therefore, using ${\rm MDS}_2$ (which we recall again that following Remark~\ref{rem:remark1} is $(T,T-B)_{\mathbb{F}_q}$ MDS code), information symbols $\{u[B-N+1],\ldots,u[k]\}$ can be recovered and cancelled from $x[i+T]$ and thus $u[i]$ can be recovered.

        \item N arbitrary erasures\newline
        {\bf Decoding information symbol $u[0]$}\newline
        First, we note that we assume that symbol $x[0]$ is one of the erased symbols otherwise decoding is trivial.\footnote{If $N=1$ it means that the only erasure is that of symbol $x[0]$ and hence decoding is done as described next for the case when symbol $x[T]$ is not erased.}
        We further differentiate between the following two cases
        \begin{itemize}
            \item Symbol $x[T]$ is erased\newline
            We note that in this case, in ${\rm MDS}_1$ we have $N-1$ erasures. Following Property~\ref{prop:pro2}, ${\rm MDS}_1$ can correct any $N-1$ erasures with maximal delay of $T$ and hence information symbol $u[0]$ can be recovered with a maximal delay of $T$.
            \item Symbol $x[T]$ is not erased\newline

            Shortening ${\rm MDS}_1$ on the first symbol results with an $(k-1+N-1,k-1)_{\mathbb{F}_q}$ MDS code with a generator matrix that is generated from $\svv{G}_1$ by deleting the first row and columns. Denoting this code as ${\rm MDS}_1^1$, its generator matrix is depicted in \eqref{eq:G1short} as the dashed \mbox{$(k-1)\times (k-1+N-1)$} lower right submatrix of $\svv{G}_1$.
    \begin{equation}
    \begin{aligned}
        \tikz[baseline=(M.west)]{%
    \node[matrix of math nodes,matrix anchor=west,left delimiter={[},right delimiter={]},ampersand replacement=\&] (M) {
        1 \& X \& \cdots \& X \& 0 \& 0 \& 0 \& \cdots \& 0 \& \alpha \& \\
        0 \& 1 \& X \& \cdots \& X \& 0 \& 0 \& \cdots \& 0 \& 0 \& \\
        0 \& 0 \& \ddots \& \ddots \& \ddots \& \ddots \& 0 \& \cdots \& 0 \& 0 \& \\
        \vdots \& \vdots \&  \& 1 \&\ddots \&\ddots \& \ddots \& \& \vdots \& X \&\\
        \vdots \& \vdots \&  \& \& \ddots \& X \& \cdots \& X \& 0 \& \vdots \&\\
        0 \& 0 \& \cdots \& \cdots \& 0 \& 1 \&  X \& \cdots \& X \& X \& \\
    };
    \coordinate (x) at (M-1-1);\coordinate (y) at (M-6-1);\draw (x) -- (y);
    \node[draw,color=blue,fit=(M-1-1)(M-6-9),inner sep=1pt,label={[xshift=0cm, yshift=0cm]$\svv{G}_1$}] {};
    \node[draw,color=blue,densely dashed,fit=(M-2-2)(M-6-9),inner sep=-1.3pt] {};
    \node[draw,color=red, dash dot,fit=(M-2-10)(M-6-10),inner sep=-1.5pt,xshift=2pt] {};
    \draw [thick, black,decorate,decoration={brace,amplitude=4pt,mirror},xshift=0pt,yshift=0.2pt](0.1,-2.2) -- (0.6,-2.2) node[black,midway,yshift=-0.6cm] {${\bf g}_0$};
    \draw [thick, black,decorate,decoration={brace,amplitude=4pt,mirror},xshift=0pt,yshift=0.2pt](0.7,-2.2) -- (5.5,-2.2) node[black,midway,yshift=-0.6cm] {${\bf g}_1,\ldots,{\bf g}_{T-1}$};
    \draw [thick, black,decorate,decoration={brace,amplitude=4pt,mirror},xshift=0pt,yshift=0.2pt](5.6,-2.2) -- (6.1,-2.2) node[black,midway,yshift=-0.6cm] {${\bf g}_T$};

    }
    \end{aligned}
    \label{eq:G1short}
    \end{equation}
    Symbols $\{x[1],\ldots,x[T-1]\}$ have up to $N-1$ erasures (since we assumed $x[0]$ is erased). Therefore, using ${\rm MDS}_1^1$, we can generate $k-1$ linear combinations of each information symbol $u[j]\in\{u[1],\ldots,u[k-1]\}$ with $u[0]$ (each linear combination consists of only $u[j]$ and $u[0]$). We denote these linear combination as $\{\tilde{u}[1],\ldots,\tilde{u}[k-1]\}$. Since $\alpha\in\mathbb{F}_{q^2}\setminus\mathbb{F}_q$, $\{{u}[1],\ldots,{u}[B-N]\}$ can be cancelled from symbol $x[T]$ using $\{\tilde{u}[1],\ldots,\tilde{u}[B-N]\}$ while it is guaranteed that $\alpha$ is not nulled out\footnote{Since all elements in ${\rm MDS}_2$ belong to $\mathbb{F}_q$, the cancellation is done by multiplying $\{\tilde{u}[1],\ldots,\tilde{u}[B-N]\}$ with coefficients $\mathbb{F}_q$.} and thus, symbol $u[0]$ can be recovered with a delay of $T$.

    Alternatively, we note that the dashed part of ${\bf g}_T$ is in the span of ${\rm MDS}_1^1$ (and further can be denoted as linear combination from $\mathbb{F}_q$ (the base field) of the symbols of ${\rm MDS}_1$). Since we have enough linear independent columns from ${\rm MDS}_1^1$, the dashed part of ${\bf g}_T$ can be cancelled while it is guaranteed that $\alpha\in\mathbb{F}_{q^2}\setminus\mathbb{F}_q$, is not nulled out and thus information symbol $u[0]$ can be recovered.
    \end{itemize}
    {\bf Decoding information symbol $u[i]\in\{u[1],\ldots,u[B-N]\}$} \newline
    We first assume by induction that $\{u[0],\ldots, u[i-1]\}$ have already been recovered by time $T+i$. We further assume that their impact on symbols $\{x[i],\ldots,x[i+T]\}$ is cancelled. After cancelling symbols $\{u[0],\ldots,u[i-1]\}$ from ${\rm MDS}_1$ we are left with a \mbox{$(k-i+N-1,k-i)_{\mathbb{F}_q}$} MDS code (which can recover any $N-1$ erasures). Equivalently, this can be viewed as shortening ${\rm MDS}_1$ by $i$ symbols. We denote it as ${\rm MDS}_1^i$.

    Since $x[i]$ is composed of (a sub group) of $\{u[0],\ldots,u[i]\}$ and we assumed that $\{u[0],\ldots,u[i-1]\}$ have been decoded correctly, if $x[i]$ is not erased, $u[i]$ can be trivially decoded.

    Therefore, assuming $x[i]$ is erased, the decoder does the following:
    \begin{itemize}
            \item In case symbol $x[i+T]$ is erased\newline
            Assuming $x[i+T]$ is erased means that there are at most $N-1$ erasures in ${\rm MDS}_1^i$. Recalling that ${\rm MDS}_1^i$ can recover any $N-1$ erasures, all information symbols can be decoded up to time $T+i$.
            \item Symbol $x[i+T]$ not erased\newline
            We note that the lower right \mbox{$k-(i+1)\times k-(i+1)+N-1$} sub-matrix of $\svv{G}_1$ is also $(k-(i+1)+N-1,k-(i+1))_{\mathbb{F}_q}$ MDS code (can be also viewed as the outcome of shortening ${\rm MDS}_1^i$ by one symbol). We denote it as ${\rm MDS}_1^{i+1}$. We may assume that symbols $\{x[i+1],\ldots,x[i+T-1]\}$ have up to $N-1$ erasures hence using ${\rm MDS}_1^{i+1}$, $k-i-1$ linear combination of $u[i]$ and  $u[j]\in\{u[i+1],\ldots,u[k-1]\}$ can be derived (each combination consists of $u[i]$ and a single element from $\{u[i+1],\ldots,u[k-1]\}$) with coefficients taken from $\mathbb{F}_q$.

            Next, $\{\tilde{u}[i+1],\ldots,\tilde{u}[B-N]\}$ are cancelled from symbol $x[i+T]$ using operations over $\mathbb{F}_q$ and, again, it is guaranteed that $\alpha\in\mathbb{F}_{q^2}\setminus\mathbb{F}_q$ is not cancelled, thus $u[i]$ can be recovered.
        \end{itemize}
    \end{itemize}

\item Decoding information symbols $\{u[B-N+1],\ldots,u[k-1]\}$\newline
    We first assume by induction that $\{u[0],\ldots, u[B-N]\}$ have already been recovered by time $T+B-N$ and cancelled from the received symbols.
    This means that we are left with ${\rm MDS}_2$. Recalling Property~\ref{prop:pro3} ${\rm MDS}_2$, is a $(k-(B-N+1)\times(n-(B-N+1)_{\mathbb{F}_q}$ MDS code which means it can correct any $B$ erasures. Recalling that $B\geq N$ it means that either a burst of $B$ erasures or arbitrary $N$ erasures can be correctly decoded at time $T+i$.
\end{itemize}

\section{Alternative proof of Theorem~\ref{thm:thm1} - analysis of the parity-check matrix}
\label{sec:parityCheck}
In \cite{KrishnanShuklaKumar2019}, a set of requirements on the parity-check matrix of $(n,k,T)_{\mathbb{F}}$-block code which is $(W,N,B)$-achievable were defined. In this section we first recall this set of requirements and than we show that the parity-check matrix of the code suggested in Section \ref{sec:codeConstruction} meet these requirements.
\subsection{Requirements on the parity-check matrix (Section II.F in \cite{KrishnanShuklaKumar2019})}
Let $i$ denote an erased coordinate. Due to the delay constraint, all symbols in time $\left[i+T+1:n-1\right]$ are unavailable to the decoder. We assume though that during the decoding of symbol $x[i]$ all symbols $x[0],\ldots,x[i-1]$ are available. Therefore we have:
\begin{align}
    \underbrace{x[0],\ldots,x[i-1]}_{\rm Known},\underbrace{x[i]}_{\rm Symbol~to~be ~recovered},\underbrace{x[i+1],\ldots,x[i+1],\ldots,x[i+T]}_{\rm All~the~non-erased~symbols~are~accessible},\underbrace{x[i+T+1],\ldots,x[n-1]}_{\rm Inaccessible~symbols}
\end{align}

Let $\mathcal{K}$ denote the set of coordinates of $\left\{0:i-1\right\}$ and $\mathcal{U}$ denote the set of coordinates of $\left\{i+T+1:n-1\right\}$. Thus when decoding $x[i]$, we assume that any $x[j],~j\in\mathcal{K}$ are known, any $x[j],~j\in\mathcal{U}$ are inaccessible and that non-erased code symbol $x[j],~j\in\left\{i+1:t+T\right\}$ are also accessible.

The effective code used to decode $x[i]$ is obtained from $\mathcal{C}$ by shortening it on the coordinates in $\mathcal{K}$ and puncturing it on coordinates in $\mathcal{U}$. Recalling Theorem~\ref{thm:thm1}, puncturing of the generator matrix is equivalent to the shortening of the parity-check matrix, and the shortening of the generator matrix is equivalent to the puncturing of the parity-check matrix.

Let $\svv{H}$ be the parity-check matrix of the code $\mathcal{C}$. For $0\leq l \leq B-N$, set
\begin{align}
    \svv{H}^{(l)}=\left[{\bf h}_0^{(l)},\ldots,{\bf    h}_{l+T}^{(l)}\right],
\end{align}
be the parity-check of the code $\mathcal{C}$ restricted to the coordinates $[0:l+T]$ (alternatively, we can say that it is the parity-check of a code which is the outcome of puncturing coordinates $[l+T+1:n]$ from the original code). This means that $\svv{H}^{(l)}$ is the derived from $\svv{H}$ by shortening coordinates $[l+T+1:n]$. Therefore, $\svv{H}^{(l)}$ is a $N+l\times T+1$ matrix.

Recalling Lemma~\ref{lemma:lemNik}, this means that in order to achieve capacity the following conditions on $\svv{H}$ and $\left\{\svv{H}^{(l)},~ 0\leq l\leq B-N\right\}$ must hold:
\begin{enumerate}
    \item {\bf Condition B1} For $0\leq\l\leq B-N$, the $l$-th column, ${\bf h}_l^{(l)}$ of $\svv{H}^{(l)}$ should be linearly independent of the set of $(B-1)$ columns
    \begin{align*}
        \left\{{\bf h}_j^{(l)},~l+1\leq j \leq l+B-1\right\}
    \end{align*}
    \item {\bf Condition R1} For $0\leq\l\leq B-N$, the $l$-th column, ${\bf h}_l^{(l)}$ of $\svv{H}^{(l)}$ should be linearly independent of the any set of $(N-1)$ columns taken from the set
    \begin{align*}
        \left\{{\bf h}_j^{(l)},~l+1\leq j \leq l+T\right\}
    \end{align*}
    \item {\bf Condition B2} For $B-N+1\leq l \leq T-N+1$, the set
    \begin{align*}
        \left\{{\bf h}_j,~l\leq j \leq l+B\right\}
    \end{align*}
    of columns of $\svv{H}$ should be linearly independent.
    \item {\bf Condition R2} Any set of $N$ columns from the set
    \begin{align*}
        \left\{{\bf h}_j,~B-N+1\leq j \leq T+B-N+1\right\}
    \end{align*}
    of columns of $\svv{H}$ should be linearly independent.
\end{enumerate}

\subsection{Rank of matrix multiplication}
In analyzing the parity matrix of the suggested code we use the following properties of the rank of matrix multiplication.
\begin{lemma}
Let $\svv{A}$ be $K\times L$ matrix and $\svv{B}$ an $L\times M$ matrix. Then
\begin{align}
    {\rm rank}(\svv{A}\svv{B})\leq \min\left( {\rm rank}(\svv{A}),{\rm rank}(\svv{B})\right)
\end{align}
\label{lem:multUppBoundByMin}
\end{lemma}
The proof of Lemma~\ref{lem:multUppBoundByMin} is given in Appendix~\ref{app:proofPfLemUpp}.

\begin{lemma}[Sylvester’s rank inequality, \cite{hohn2013elementary}]
Let $\svv{A}$ be $K\times L$ matrix and $\svv{B}$ an $L\times M$ matrix. Then
\begin{align}
    {\rm rank}(\svv{A}\svv{B})\geq {\rm rank}(\svv{A})+{\rm rank}(\svv{B})-L
\end{align}
\label{prop:SylvesterTheorem}
\end{lemma}
\begin{corollary}
If $\svv{A}$ is a square ($L\times L$) full rank matrix, Sylvester’s rank inequality means that
\begin{align}
    {\rm rank}(\svv{A}\svv{B})= {\rm rank}(\svv{B}).
\end{align}
If $\svv{B}$ is a ($L\times L$) full rank matrix, from Sylvester’s rank inequality we have
\begin{align}
{\rm rank}(\svv{A}\svv{B})= {\rm rank}(\svv{A}).
\end{align}
\label{col:colSylFullRank}
\end{corollary}
\subsection{Analysis of the parity-check matrix of the suggested block code}
As mentioned in Remark~\ref{rem:remark3}, the suggested code can be transformed to a systematic block code. The generator matrix of the systematic code can be denoted as
\begin{align}
    \svv{\tilde{G}}=\svv{M}^{-1}\svv{G}
    \label{eq:tildeG}
\end{align}
where $\svv{M}$ and $\svv{G}$ are defined in \eqref{eq:matM} and \eqref{eq:matG}.

Since $\svv{G}$ is generated from $\svv{G}'$ (defined in~\eqref{eq:matGprime}) by replacing the $(B-N+1)\times (B-N+1)$ upper right matrix with $\alpha\cdot\svv{I}_{B-N+1}$ where $\alpha\in\mathbb{F}_{q^2}\setminus\mathbb{F}_q$ and recalling that $\svv{G}'=\svv{M}\svv{G}''$ (where $\svv{G}''$ is a systematic matrix defined in \eqref{eq:matGDoublePrime}) we have that
\begin{equation}
    \renewcommand\arraystretch{1.8}
    \begin{aligned}
    \svv{\tilde{G}}=&
    \tikz[baseline=(M.west)]{%
    \node[matrix of math nodes,matrix anchor=west,left delimiter={[},right delimiter={]},ampersand replacement=\& ] (M) {
    1 \& 0 \& \cdots \& \cdots \& \cdots \& \cdots \& 0 \&  Y \& \cdots \& Y \& f_{1,1}(\alpha) \& \cdots \& \cdots \& f_{1,B-N+1}(\alpha) \\
    0 \& 1 \& 0 \& \cdots \& \cdots \& \cdots \& 0 \& Y \& \cdots \& Y \&  Y_{\alpha} \& f_{2,2}(\alpha) \&  \cdots \& f_{2,B-N+1}(\alpha) \\
    0 \& 0 \& \ddots \& \cdots \& \vdots \& \vdots \& 0 \& Y \& \cdots \& Y \& \vdots \& \ddots \& \ddots \& \vdots     \\
    \vdots \& \vdots \&  \& \ddots \&\vdots \& \vdots \& 0 \& Y \& \cdots \& Y  \&  Y_{\alpha} \& \cdots \&  Y_{\alpha} \& f_{B-N+1,B-N+1}(\alpha)      \\
    \vdots \& \vdots \&  \& \& \ddots \&  \vdots \& 0 \& Y \& \cdots \& Y  \&  Y \& \cdots \& Y \& Y \\
    0 \& 0 \& \cdots \& \cdots \& 0 \& 1 \& 0 \&  Y \& \cdots \& Y \& Y \& \cdots \& Y \& Y \\
    0 \& 0 \& \cdots\& \cdots \&  \cdots  \& 0 \& 1 \& Y \& \cdots \& Y  \& Y \& \cdots \& Y  \& Y \\
    };
    \draw[decorate,decoration={brace,raise=4mm,amplitude=10pt,mirror}](M-7-1.west) -- node [below of=M,below=-3mm] {$k$} (M-7-7.east);
    \draw[decorate,decoration={brace,raise=4mm,amplitude=10pt,mirror}](M-7-8.west) -- node [below of=M,below=-3mm] {$N-1$} (M-7-10.east);
    \draw[decorate,decoration={brace,raise=4mm,amplitude=10pt,mirror}](M-7-11.west) -- node [below of=M,below=-3mm] {$B+N-1$} (M-7-14.east);
    }.
    \end{aligned}
    \label{eq:matGhat}
\end{equation}
where $f_{i,j}(\alpha)$ is some function of $\alpha$ (hence it contains elements from $\mathbb{F}_{q^2}\setminus\mathbb{F}_q$), and $Y_{\alpha}$ stands for elements from the base field ($\mathbb{F}_q$) which are now different from the original elements in $\svv{P}''$. More details on the generation of $\svv{\tilde{G}}$ and each of its elements are given in Appendix~\ref{app:explicitDescOfTildeG}.

We note the following property
\begin{property}
For any $i\in\{0,\ldots,B-N+1\}$, $f_{i,i}(\alpha)\neq 0$.
\label{prop:f_ii_not_zero}
\end{property}
This property holds since $f_{i,i}(\alpha)$ is the outcome of multiplying the $i$'th row in $\svv{M}^{-1}$ with the $k+i$ column of $\svv{G}$. Since $\svv{M}^{-1}_{i,i}=1$ and this is the only element which multiplies $\alpha$ it follows that $f_{i,i}(\alpha)\neq 0$ (summing $\alpha$ with elements from $\mathbb{F}_{q}$ cannot null it).

Therefore, the parity-check matrix can be written as
\begin{equation}
\renewcommand\arraystretch{2}
\begin{aligned}
\svv{H}&=\left[-\svv{\tilde{P}}^T\given[\Big]\svv{I}_{B\times B}\right] \\
&=\tikz[baseline=(M.west)]{%
    \node[matrix of math nodes,matrix anchor=west,left delimiter={[},right delimiter={]},ampersand replacement=\& ] (M) {
    -Y \& \cdots \& \cdots \&  \cdots \& -Y \&  \cdots \& -Y \& 1 \& 0 \& \cdots \& \cdots \& \cdots \& \cdots \& 0   \\
\vdots \& \vdots \& \vdots \&  \vdots \&  \vdots \&  \vdots \& \vdots \& 0 \& 1 \& 0 \& \cdots \& \cdots \& \cdots \& 0 \\
-Y \& \cdots \& \cdots \&  \cdots \&  -Y \&  \cdots \&  -Y \& 0 \& 0 \& \ddots \& \cdots \& \vdots \& \vdots \& 0    \\
f_{1,1}(\alpha) \& -Y_{\alpha} \& \cdots \& -Y_{\alpha} \& -Y \& \cdots \& -Y \& \vdots \& \vdots \&  \& 1 \& \ddots \&  \vdots \& 0    \\
f_{1,2}(\alpha) \& f_{2,2}(\alpha) \& \ddots \& \vdots \& \vdots \& \cdots \& -Y \& \vdots \& \vdots \&  \& 0 \& \ddots \& \vdots \& 0    \\
\vdots \& \vdots \& \ddots \& -Y_{\alpha} \&\vdots \&\vdots  \& -Y \& \vdots \& \vdots \&  \& \& 0 \&  \ddots \& 0    \\
f_{1,B-N+1}(\alpha) \& f_{2,B-N+1}(\alpha) \& \cdots \& f_{B-N+1,B-N+1}(\alpha) \&-Y \&\cdots  \& -Y \& 0 \& 0 \& \cdots \& \cdots \& 0 \& 0 \& 1 \\
};
    \node[draw,color=blue,fit=(M-1-11)(M-4-1),inner sep=1pt,label=below:\textcolor{blue}{}] {};
    \node[draw,color=red,densely dashed,fit=(M-5-1)(M-1-12),inner sep=3pt,label=below:\textcolor{red}{}] {};
    \node[draw,color=green,dashdotted,fit=(M-7-1)(M-1-14),inner sep=0pt,label=below:\textcolor{green}{}] {};
    \draw[decorate,decoration={brace,raise=4mm,amplitude=10pt,mirror}](M-7-1.west) -- node [below of=M,below=-3mm] {$B-N+1$} (M-7-4.east);
    \draw[decorate,decoration={brace,raise=4mm,amplitude=10pt,mirror}](M-7-5.west) -- node [below of=M,below=-3mm] {$T-B$} (M-7-7.east);
    \draw[decorate,decoration={brace,raise=4mm,amplitude=10pt,mirror}](M-7-8.west) -- node [below of=M,below=-3mm] {$B$} (M-7-14.east);
    \draw[decorate,decoration={brace,raise=14mm,amplitude=10pt,mirror}](M-1-1.north) -- node [left of=M,left=7mm] {$N-1$} (M-3-1.south);
    \draw[decorate,decoration={brace,raise=14mm,amplitude=10pt,mirror}](M-4-1.north) -- node [left of=M,left=7mm] {$B-N+1$} (M-7-1.south);
    
    }.
\end{aligned}
\label{eq:H}
\end{equation}
We note that several examples of $\svv{H}^{(l)}$ are marked (each in different color and different type of frame) in $\eqref{eq:H}$. 

\begin{enumerate}
    \item {\bf Condition B1}
\end{enumerate}

Condition B1 requires that ${\bf h}_l^{(l)}$ is linearly independent of the set of $(B-1)$ columns which follows it. Therefore, we first limit $\svv{{H}}^{(l)}$ to contain a maximum of $B$ columns starting for its $l$'th column. Further, we note the following
\begin{itemize}
    \item If $B+l<k$. We denote the the group of $B$ columns starting from the $l$'th column of $\svv{{H}}^{(l)}$ as $\svv{\hat{H}}^{(l)}=\svv{{H}}^{(l)}_{:,l:l+B}$. It follows the dimensions of $\svv{\hat{H}}^{(l)}$ are $N+l\times B$.
    \item If $B+l\geq k$. We note than when $B+l\geq k$, the group of $B$ columns starting from the $l$'th column of $\svv{{H}}^{(l)}$ contains $B+l-k$ vectors from $\svv{I}_{B\times B}$ (the unit matrix). These $B+l-k$ are part of the basis that spans this group of $B$ vectors. Hence, its suffices to analyze this group without the top $B+l-k$ rows and last $B+l-k$ columns. Therefore, we denote by $\svv{\hat{H}}_{(l)}$ the group of $B$ vectors starting from $l$'th column without the top $B+l-k$ rows and last $B+l-k$. Hence $\svv{\hat{H}}_{(l)}=\svv{H}^{(l)}_{B+l-k+1:N+l+1,l+1:k}$. It follows that the dimensions of $\svv{\hat{H}}_{(l)}$ are $N+(k-B)\times k-l$.
\end{itemize}
Some examples of $\svv{\hat{H}}^{(l)}$ (where $l=0$ and $l=k-B$) and for $\svv{\hat{H}}_{(l)}$ (where $l=B-N)$ are given in \eqref{eq:hHat} where each $\svv{\hat{H}}^{(l)}$ is denoted with different color and different type of frame.

\begin{equation}
\renewcommand\arraystretch{2}
\begin{aligned}
\svv{H}&=\left[-\svv{\tilde{P}}^T\given[\Big]\svv{I}_{B\times B}\right] \\
&=\tikz[baseline=(M.west)]{%
    \node[matrix of math nodes,matrix anchor=west,left delimiter={[},right delimiter={]},ampersand replacement=\& ] (M) {
    -Y \& \cdots \& \cdots \&  \cdots \& -Y \&  \cdots \& -Y \& 1 \& 0 \& \cdots \& \cdots \& \cdots \& \cdots \& 0   \\
\vdots \& \vdots \& \vdots \&  \vdots \&  \vdots \&  \vdots \& \vdots \& 0 \& 1 \& 0 \& \cdots \& \cdots \& \cdots \& 0 \\
-Y \& \cdots \& \cdots \&  \cdots \&  -Y \&  \cdots \&  -Y \& 0 \& 0 \& \ddots \& \cdots \& \vdots \& \vdots \& 0    \\
f_{1,1}(\alpha) \& -Y_{\alpha} \& \cdots \& -Y_{\alpha} \& -Y \& \cdots \& -Y \& \vdots \& \vdots \&  \& 1 \& \ddots \&  \vdots \& 0    \\
f_{1,2}(\alpha) \& f_{2,2}(\alpha) \& \ddots \& \vdots \& \vdots \& \cdots \& -Y \& \vdots \& \vdots \&  \& 0 \& \ddots \& \vdots \& 0    \\
\vdots \& \vdots \& \ddots \& -Y_{\alpha} \&\vdots \&\vdots  \& -Y \& \vdots \& \vdots \&  \& \& 0 \&  \ddots \& 0    \\
f_{1,B-N+1}(\alpha) \& f_{2,B-N+1}(\alpha) \& \cdots \& f_{B-N+1,B-N+1}(\alpha) \&-Y \&\cdots  \& -Y \& 0 \& 0 \& \cdots \& \cdots \& 0 \& 0 \& 1 \\
};
    \node[draw,color=blue,fit=(M-1-5)(M-4-1),inner sep=1pt,label=below:\textcolor{blue}{$\svv{\hat{H}}^{(0)}$}] {};
    \node[draw,color=red,densely dashed,fit=(M-5-2)(M-1-7),inner sep=3pt,label=below:\textcolor{red}{$\svv{\hat{H}}^{(k-B)}$}] {};
    \node[draw,color=green,dashdotted,fit=(M-7-4)(M-5-7),inner sep=0pt,label=below:\textcolor{green}{$\svv{\hat{H}}_{(B-N)}$}] {};
    \draw[decorate,decoration={brace,raise=4mm,amplitude=10pt,mirror}](M-7-1.west) -- node [below of=M,below=-3mm] {$B-N+1$} (M-7-4.east);
    \draw[decorate,decoration={brace,raise=4mm,amplitude=10pt,mirror}](M-7-5.west) -- node [below of=M,below=-3mm] {$T-B$} (M-7-7.east);
    \draw[decorate,decoration={brace,raise=14mm,amplitude=10pt,mirror}](M-1-1.north) -- node [left of=M,left=7mm] {$N-1$} (M-3-1.south);
    \draw[decorate,decoration={brace,raise=14mm,amplitude=10pt,mirror}](M-4-1.north) -- node [left of=M,left=7mm] {$B-N+1$} (M-7-1.south);
    }
    \end{aligned}
    \label{eq:hHat}
\end{equation}

Before showing condition B1 holds, we show the following properties which will be used in the proof. Denote by $\svv{G}_{T-B,B}$ the lower right $(T-B\times B)$ matrix of $\svv{G}$ and with $\svv{M}^{-1}_{B,T-B}$ as the upper right $(B\times T-B)$ matrix of $\svv{M}^{-1}$.

\begin{property}
$\svv{G}_{T-B,B}$ has rank of $\min(T-B,B)$. Further, any sub-matrix formed by taking $w$ columns of  $\svv{G}_{T-B,B}$ has a rank of $\min(T-B,w)$.
\label{prop:prop4}
\end{property}
\begin{proof}
See Appendix~\ref{sec:app_prop4}
\end{proof}

\begin{property}
The first $B-N+1$ rows of $\svv{M}^{-1}_{B,T-B}$ are a linear function of the last $N-1$ rows.
\label{prop:prop5}
\end{property}
\begin{proof}
See Appendix~\ref{sec:app_prop5}.
\end{proof}
    \begin{itemize}
        \item $B+l< k$: analyzing $\svv{\hat{H}}^{(l)}$.
    \end{itemize}
    Recalling that $\svv{\hat{H}}^{(l)}$ has $N+l$ rows and $B$ columns, an example of $\svv{\hat{H}}^{(l)}$ (marked in solid) is
        \begin{equation}
    \begin{aligned}
    \svv{\hat{H}}^{(l)}& = \svv{{H}}^{(l)}_{1:N+l,l+1:l+B} \nonumber \\
    &=
        \tikz[baseline=(M.west)]{%
    \node[matrix of math nodes,matrix anchor=west,left delimiter={[},right delimiter={]},ampersand replacement=\& ] (M) {
        -Y \& \cdots \& \cdots \&  \cdots \&  \cdots \&  \cdots \&  -Y \\
    \vdots \& \vdots \& \vdots \&  \vdots \&  \vdots \&  \vdots \&  \vdots \\
    -Y \& \cdots \& \cdots \&  \cdots \&  \cdots \&  \cdots \&  -Y \\
    -Y_{\alpha} \& \cdots \& \cdots \&  -Y_{\alpha} \&  -Y \&  \cdots \&  -Y \\
    \vdots \& \vdots \& \vdots \&  \vdots \&  \vdots \&  \vdots \&  \vdots \\
    f_{l+1,l+1}(\alpha) \& -Y_{\alpha} \& \cdots \& -Y_{\alpha} \& -Y \& \cdots \& -Y \\
        };
    \node[draw,color=blue,fit=(M-6-1)(M-1-6),inner sep=2pt,label={[xshift=-1cm, yshift=0cm]}] {};
    \node[draw,color=blue,densely dashed,fit=(M-1-5)(M-5-6),inner sep=-2pt,label={[xshift=-1cm, yshift=0cm]}] {};
    \draw[decorate,decoration={brace,raise=4mm,amplitude=10pt,mirror}](M-6-1.west) -- node [below of=M,below=-3mm] {$B-N+1-l$} (M-6-4.east);
    \draw[decorate,decoration={brace,raise=4mm,amplitude=10pt,mirror}](M-6-5.west) -- node [below of=M,below=-3mm] {$N+l-1$} (M-6-6.east);
    \draw[decorate,decoration={brace,raise=12mm,amplitude=10pt,mirror}](M-6-1.west) -- node [below of=M,below=6mm] {$B$} (M-6-6.east);
    \draw[decorate,decoration={brace,raise=13mm,amplitude=10pt,mirror}](M-1-1.north) -- node [left of=M,left=6mm] {$N+l$} (M-6-1.south);
    }
    \end{aligned}.
    \label{eq:HlCau}
\end{equation}

We start with analyzing the rank of $\svv{\hat{H}}^{(l)}$. First, we note that since we want to show condition B1 for $0\leq l \leq B-N$ it follows that $N+l\leq B$ thus the maximal rank of $\svv{\hat{H}}^{(l)}$ is $N+l$.

We note that when $B+l\leq k$, each $\svv{\hat{H}}^{(l)}$ contains a $(N+l-1\times N+l-1)$ sub-matrix of $\svv{P}''$ marked in dashed in \eqref{eq:HlCau}. Following Corollary~\ref{col:CauchyForRectangular}, each sub-matrix of $\svv{P}''$ has a full rank thus the rank of this sub-matrix is $N+l-1$. This means we can transform $\svv{\hat{H}}^{(l)}$ to an equivalent row echelon form of
        \begin{equation}
    \left[\begin{tabular}{ c c c c c c c}
        \multicolumn{4}{c}{${\bf 0}_{N+l-1\times B-(N+l-1)}$} & \multicolumn{3}{c}{$\svv{I}_{N+l-1}$}  \\
         $\tilde{f}_{l+1,l+1}(\alpha)$ & $-\tilde{Y}_{\alpha}$ & $\cdots$ & $-\tilde{Y}_{\alpha}$ & $-\tilde{Y}$ & $\cdots$ & $-\tilde{Y}$
    \end{tabular}\right],
\end{equation}
where $\tilde{f}_{l+1,l+1}(\alpha)$ ,each $-\tilde{Y}_{\alpha}$ and each $-\tilde{Y}$ are the outcome of the column operations performed on $\svv{\hat{H}}^{(l)}$ (over $\mathbb{F}_q$). Following Property~\ref{prop:f_ii_not_zero} we have that $f_{l+1,l+1}(\alpha)\neq 0$ and since all column operations were performed over $\mathbb{F}_q$ it follows that $\tilde{f}_{l+1,l+1}(\alpha)\neq 0$. Thus, we have that 
\begin{align}
    {\rm rank}(\svv{\hat{H}}^{(l)})= N+l.
    \label{eq:rankH^l}
\end{align}

Proving condition B1 is equivalent to prove that column ${\bf \hat{h}}_0^{(l)}$ (the first column in $\svv{\hat{H}}^{(l)}$) is linearly independent of the rest of the $B-1$ columns of $\svv{\hat{H}}^{(l)}$ which we denote as $\svv{\hat{H}}^{(l)}_{2:B}$. We note that $\svv{\hat{H}}^{(l)}$ is generated by taking the transpose of multiplying $\svv{M}^{-1}_{(l+1:B+l,l+1:k)}$ with $\svv{G}_{(l+1:k,k+1:k+N+l)}$ which we denote as $\svv{G}^{(l)}$
Hence,
\begin{align}
    \svv{\hat{H}}^{(l)}=\left(\svv{M}^{-1}_{(l+1:B+l,l+1:k)}\times\svv{G}^{(l)}\right)^T.
\end{align}

In \eqref{eq:Goption1} below we show examples of $\svv{G}^{(l)}$ when we assume $B+l\leq k$ for all $0\leq l \leq B-N$.
\begin{equation}
    \begin{aligned}
    \svv{G}=
        \tikz[baseline=(M.west)]{%
    \node[matrix of math nodes,matrix anchor=west,left delimiter={[},right delimiter={]},ampersand replacement=\& ] (M) {
        1 \& X \& \cdots \& X \& 0 \& 0 \& 0 \& \cdots \& 0 \& \alpha \& \cdots \& 0 \\
        0 \& 1 \& X \& \cdots \& X \& 0 \& 0 \& \cdots \& 0 \& 0 \& \ddots \& 0 \\
        0 \& 0 \& \ddots \& \ddots \& \ddots \& \ddots \& 0 \& \cdots \& 0 \&  0 \& \cdots \&  \alpha \\
        \vdots \& \vdots \&  \& 1 \&\ddots \&\ddots \& \ddots \& \& \vdots \& X \& \cdots \& X \\
        \vdots \& \vdots \&  \& \& \ddots \& X \& \cdots \& X \& 0 \& \vdots \& \& \vdots \\
        0 \& 0 \& \cdots \& \cdots \& 0 \& 1 \&  X \& \cdots \& X \& X \& \cdots \& X \\
        };
    \node[draw,color=blue,fit=(M-1-7)(M-6-10),inner sep=2pt,label={[xshift=0cm, yshift=0cm]\textcolor{blue}{$\svv{G}^{(0)}$}}] {};
    \node[draw,color=red,fit=(M-2-7)(M-6-11),inner sep=1pt,label={[xshift=0cm, yshift=0cm]\textcolor{red}{$\svv{G}^{(1)}$}}] {};
    \node[draw,color=green,fit=(M-3-7)(M-6-12),inner sep=0pt,label={[xshift=0cm, yshift=0cm]}] {};
    \draw[decorate,decoration={brace,raise=7mm,amplitude=10pt}](M-1-12.north) -- node [right of=M,right=3mm] {$B-N+1$} (M-3-12.south);
    \draw[decorate,decoration={brace,raise=7mm,amplitude=10pt}](M-4-12.north) -- node [right of=M,right=3mm] {$T-B$} (M-6-12.south);
    \draw[decorate,decoration={brace,raise=4mm,amplitude=10pt,mirror}](M-6-1.west) -- node [below of=M,below=-3mm] {$k$} (M-6-6.east);
    \draw[decorate,decoration={brace,raise=4mm,amplitude=10pt,mirror}](M-6-7.west) -- node [below of=M,below=-3mm] {$N-1$} (M-6-9.east);
    \draw[decorate,decoration={brace,raise=4mm,amplitude=10pt,mirror}](M-6-10.west) -- node [below of=M,below=-3mm] {$B-N+1$} (M-6-12.east);
    }
    \end{aligned}.
    \label{eq:Goption1}
\end{equation}

We note that since in generating $\svv{G}$ we replaced the upper right $(B-N+1)\times(B-N+1)$ matrix of $\svv{G}'$ with $\alpha \svv{I}_{B-N+1}$ it means that there are only $T-B+1$ non-zero rows  in $\svv{G}^{(l)}_{(l+2:k,:)}$ (the first row and the last $T-B$ rows).

Therefore, denoting with $\svv{G}^{(l)}_{T-B,:}$ the last $T-B$ rows of $\svv{G}^{(l)}$, we have
\begin{align}
    \svv{\hat{H}}^{(l)}_{2:B}=\left(\svv{M}^{-1}_{(l+2:B+l,2+B-N:k)}\times \svv{G}^{(l)}_{T-B,:}\right)^T,
\end{align}
where $\svv{M}^{-1}_{(l+2:B+l,2+B-N:k)}$ is acquired by taking the $B-1\times T-B$ lower right matrix of $\svv{M}^{-1}_{(l+2:B+l,1:k)}$. Several examples of $\svv{M}^{-1}_{(l+2:B+l,2+B-N:k)}$ and $\svv{M}^{-1}_{(l+2:B+l,1:k)}$ for different values of $l$ (where the full matrix is marked in solid and the partial matrix is marked in dashed) are shown in \eqref{eq:MinvExm}.


\begin{equation}
    \renewcommand\arraystretch{1.8}
    \svv{M}^{-1}=
    \tikz[baseline=(M.west)]{%
    \node[matrix of math nodes,matrix anchor=west,left delimiter={[},right delimiter={]},ampersand replacement=\& ] (M) {
    1 \& Y'' \& \cdots \& Y'' \& Y'' \& \cdots \& \cdots \& Y'' \\
    0 \& 1 \& Y'' \& \cdots \& Y'' \& Y'' \& \cdots \& Y'' \\
    \vdots \& \cdots \& \ddots \& \ddots \& \ddots \& \ddots \& \ddots \& \vdots\\
    0 \& \cdots \& 0 \& 1 \& Y'' \& \cdots \& Y'' \& Y'' \\
    0 \& \cdots \& 0 \& 0 \& 1 \& Y'' \& \cdots \& Y'' \\
    \vdots \& \cdots \& \vdots \& \vdots \& \vdots \& \ddots \& \ddots \& Y'' \\
    0 \& \cdots \& 0 \& 0 \& 0 \& 0 \& 1 \& Y'' \\
    0 \& \cdots \& 0 \& 0 \& 0 \& 0 \& 0 \& 1 \\
};
    \node[draw,color=blue,fit=(M-1-1)(M-5-8),inner sep=2pt,label={[xshift=-1cm, yshift=0cm]}] {};
    \node[draw,color=blue,densely dashed,fit=(M-2-4)(M-5-8),inner sep=-1pt,label={[xshift=-1cm, yshift=0cm]}] {};
    \node[draw,color=black,fit=(M-4-4)(M-5-8),inner sep=-1pt,label={[xshift=-1cm, yshift=0cm]}] {};
    \draw[decorate,decoration={brace,raise=5mm,amplitude=10pt,mirror}](M-1-1.north) -- node [left of=M,left=-1mm] {$B$} (M-5-1.south);
    \draw[decorate,decoration={brace,raise=3mm,amplitude=10pt,mirror}](M-8-4.west) -- node [below of=M,below=-3mm] {$T-B$} (M-8-8.east);
    \draw[decorate,decoration={brace,raise=7mm,amplitude=6pt}](M-4-8.north) -- node [right of=M,right=-1mm] {$N-1$} (M-5-8.south);

    },~ 
    \svv{M}^{-1}=
    \tikz[baseline=(M.west)]{%
    \node[matrix of math nodes,matrix anchor=west,left delimiter={[},right delimiter={]},ampersand replacement=\& ] (M) {
    1 \& Y'' \& \cdots \& Y'' \& Y'' \& \cdots \& \cdots \& Y'' \\
    0 \& 1 \& Y'' \& \cdots \& Y'' \& Y'' \& \cdots \& Y'' \\
    \vdots \& \cdots \& \ddots \& \ddots \& \ddots \& \ddots \& \ddots \& \vdots\\
    0 \& \cdots \& 0 \& 1 \& Y'' \& \cdots \& Y'' \& Y'' \\
    0 \& \cdots \& 0 \& 0 \& 1 \& Y'' \& \cdots \& Y'' \\
    \vdots \& \cdots \& \vdots \& \vdots \& \vdots \& \ddots \& \ddots \& Y'' \\
    0 \& \cdots \& 0 \& 0 \& 0 \& 0 \& 1 \& Y'' \\
    0 \& \cdots \& 0 \& 0 \& 0 \& 0 \& 0 \& 1 \\
};
    \node[draw,color=red,fit=(M-2-2)(M-6-8),inner sep=1pt,label={[xshift=-1cm, yshift=0cm]}] {};
    \node[draw,color=red,densely dashed,fit=(M-3-4)(M-6-8),inner sep=-1pt,label={[xshift=-1cm, yshift=0cm]}] {};
    \node[draw,color=black,fit=(M-4-4)(M-5-8),inner sep=-1pt,label={[xshift=-1cm, yshift=0cm]}] {};
    \draw[decorate,decoration={brace,raise=5mm,amplitude=10pt,mirror}](M-2-1.north) -- node [left of=M,left=-1mm] {$B$} (M-6-1.south);
    \draw[decorate,decoration={brace,raise=3mm,amplitude=10pt,mirror}](M-8-4.west) -- node [below of=M,below=-3mm] {$T-B$} (M-8-8.east);
    \draw[decorate,decoration={brace,raise=7mm,amplitude=6pt}](M-4-8.north) -- node [right of=M,right=-1mm] {$N-1$} (M-5-8.south);

    },
    \label{eq:MinvExm}
\end{equation}

Noting that $\svv{M}^{-1}_{(l+2:B+l,2+B-N:k)}$ contains $B-N-l$ rows from the first $B-N+1$ rows of $\svv{M}^{-1}_{B,T-B}$ (depicted in \eqref{eq:invM_TmBCol}), and  recalling Property~\ref{prop:prop5}, it follows that these rows are a linear function of the last $N-1$ rows of $\svv{M}^{-1}_{B,T-B}$ (which are also a part of $\svv{M}^{-1}_{(l+2:B+l,2+B-N:k)}$). Further, the lower $N-1+l$ rows of $\svv{M}^{-1}_{(l+1:B+l,2+B-N:k)}$ are an upper triangular matrix therefore,
\begin{align}
    {\rm rank}(\svv{M}^{-1}_{(l+1:B+l,2+B-N:k)})=N-1+l.
    \label{eq:rankM_min1_first}
\end{align}

Since taking the last $T-B$ rows of $\svv{G}^{(l)}$ results with a matrix which is a sub-matrix of $\svv{G}_{T-B,B}$ (by taking only $N+l$ columns out of the $B$ columns of $\svv{G}_{T-B,B}$), recalling Property~\ref{prop:prop4}, it follows that
\begin{align}
    {\rm rank}(\svv{G}^{(l)}_{T-B,:})=\min(T-B,N+l).
    \label{eq:rankG_first}
\end{align}

Therefore, following Lemma~\ref{lem:multUppBoundByMin}, \eqref{eq:rankM_min1_first} and \eqref{eq:rankG_first} we have
\begin{align}
    {\rm rank}(\svv{\hat{H}}^{(l)}_{2:B})\leq & \min\left({\rm rank}(\svv{M}^{-1}_{(l+2:B+l,2+B-N:k)} ),{\rm rank}(\svv{G}^{(l)}_{T-B,:}) \right) \nonumber \\
    & \min (T-B,N-1+l)
\end{align}
and since we assume
\begin{align}
    B+l&\leq k \nonumber \\
    &= T-N+1,
\end{align}
we have $N-1+l\leq T-B$. Thus,
\begin{align}
    {\rm rank}(\svv{\hat{H}}^{(l)}_{2:B})\leq N-1+l
    \label{eq:rankH^lB-1}
\end{align}

Therefore, since ${\rm rank}(\svv{\hat{H}}^{(l)})=N+l$ and ${\rm rank}(\svv{\hat{H}}^{(l)}_{2:B})\leq N-1+l$, it follows that ${\bf {\hat h}}^{(l)}_0$ is linearly independent with the following $B-1$ columns.


\begin{itemize}
\item $B+l\geq k$: analyzing $\svv{\hat{H}}_{(l)}$.
\end{itemize}
Recalling that $\svv{\hat{H}}_{(l)}$ has $N+k-B$ rows and $k-l$ columns, an example of $\svv{\hat{H}}_{(l)}$ is
    \begin{equation}
    \begin{aligned}
    \svv{\hat{H}}_{(l)} &= \svv{{H}}^{(l)}_{B+l+k+1:N+l,l+1:k} \nonumber \\
    & = 
        \tikz[baseline=(M.west)]{%
    \node[matrix of math nodes,matrix anchor=west,left delimiter={[},right delimiter={]},ampersand replacement=\& ] (M) {
        -Y \& \cdots \& \cdots \&  \cdots \&  \cdots \&  \cdots \&  -Y \\
    \vdots \& \vdots \& \vdots \&  \vdots \&  \vdots \&  \vdots \&  \vdots \\
    -Y \& \cdots \& \cdots \&  \cdots \&  \cdots \&  \cdots \&  -Y \\
    -Y_{\alpha} \& \cdots \& \cdots \&  -Y_{\alpha} \&  -Y \&  \cdots \&  -Y \\
    \vdots \& \vdots \& \vdots \&  \vdots \&  \vdots \&  \vdots \&  \vdots \\
    f_{l+1,l+1}(\alpha) \& -Y_{\alpha} \& \cdots \& -Y_{\alpha} \& -Y \& \cdots \& -Y \\
        };
    \node[draw,color=blue,fit=(M-6-1)(M-3-7),inner sep=2pt,label={[xshift=-1cm, yshift=0cm]}] {};
    \node[draw,color=blue,densely dashed,fit=(M-5-5)(M-3-7),inner sep=0pt,label={[xshift=-1cm, yshift=0cm]}] {};
    \draw[decorate,decoration={brace,raise=4mm,amplitude=10pt,mirror}](M-6-1.west) -- node [below of=M,below=-3mm] {$B-N+1-l$} (M-6-4.east);
    \draw[decorate,decoration={brace,raise=4mm,amplitude=10pt,mirror}](M-6-5.west) -- node [below of=M,below=-3mm] {$T-B$} (M-6-7.east);
    \draw[decorate,decoration={brace,raise=12mm,amplitude=10pt,mirror}](M-6-1.west) -- node [below of=M,below=6mm] {$k-l$} (M-6-7.east);
    \draw[decorate,decoration={brace,raise=13mm,amplitude=10pt,mirror}](M-1-1.north) -- node [left of=M,left=6mm] {$B+l-k$} (M-2-1.south);
    \draw[decorate,decoration={brace,raise=13mm,amplitude=10pt,mirror}](M-3-1.north) -- node [left of=M,left=6mm] {$N+k-B$} (M-6-1.south);
    \draw[decorate,decoration={brace,raise=33mm,amplitude=10pt,mirror}](M-1-1.north) -- node [left of=M,left=26mm] {$N+l$} (M-6-1.south);
    }
    \end{aligned}.
    \label{eq:HlCau_2}
\end{equation}

We first note that each $\svv{\hat{H}}_{(l)}$ contains a $(N+k-B-1\times T-B)$ sub-matrix of $\svv{P}''$ (marked in dashed in~\eqref{eq:HlCau_2}). We note that
  \begin{align}
      N+k-B-1&=N+(T-N+1)-B-1 \nonumber \\
      &=T-B
      \label{eq:eq1}
\end{align}
Hence, the rank of this matrix is $T-B$. 

This means we can transform $\svv{\hat{H}}_{(l)}$ to an equivalent row echelon form of
        \begin{equation}
    \svv{\hat{H}}_{(l)}=\left[\begin{tabular}{ c c c c c c c}
        \multicolumn{4}{c}{${\bf 0}_{T-B\times B-N+1-l}$} & \multicolumn{3}{c}{$\svv{I}_{T-B}$}  \\
         $\tilde{f}_{l+1,l+1}(\alpha)$ & $-\tilde{Y}_{\alpha}$ & $\cdots$ & $-\tilde{Y}_{\alpha}$ & $-\tilde{Y}$ & $\cdots$ & $-\tilde{Y}$
    \end{tabular}\right]
\end{equation}
where $\tilde{f}_{l+1,l+1}(\alpha)$ ,each $-\tilde{Y}_{\alpha}$ and each $-\tilde{Y}$ are the outcome of the column operations performed on $\svv{\hat{H}}_{(l)}$. Following Property~\ref{prop:f_ii_not_zero} we have that $f_{l+1,l+1}(\alpha)\neq 0$ and since all column operations were performed over $\mathbb{F}_q$ it follows that $\tilde{f}_{l+1,l+1}(\alpha)\neq 0$. Thus, we have that 

\begin{align}
    {\rm rank}(\svv{\hat{H}}_{(l)})=T-B+1.
\end{align}


Proving condition B1 is equivalent to prove that column ${\bf \hat{h}}_{(l),0}$ (the first column in $\svv{\hat{H}}_{(l)}$) is linearly independent of the rest of the $k-1-l$ columns of $\svv{\hat{H}}_{(l)}$ which we denote as $\svv{\hat{H}}_{(l),2:k-l}$. We note that $\svv{\hat{H}}_{(l)}$ is generated by taking the transpose of multiplying $\svv{M}^{-1}_{l+1:k,l+1:k}$ with $\svv{G}_{l+1+l:k,k+1:k+N+l}$ which we denoted as $\svv{G}_{(l)}$.


Hence,
\begin{align}
    \svv{\hat{H}}_{(l)}=\left(\svv{M}^{-1}_{(l+1:k,l+1:k)}\times\svv{G}_{(l)}\right)^T.
\end{align}
In \eqref{eq:Goption2} below we show examples of $\svv{G}_{(l)}$ when we assume $B+l > k$ for all $0\leq l \leq B-N$.

\begin{equation}
    \begin{aligned}
    \svv{G}=
        \tikz[baseline=(M.west)]{%
    \node[matrix of math nodes,matrix anchor=west,left delimiter={[},right delimiter={]},ampersand replacement=\& ] (M) {
        1 \& X \& \cdots \& X \& 0 \& 0 \& 0 \& \cdots \& 0 \& \alpha \& \cdots \& 0 \\
        0 \& 1 \& X \& \cdots \& X \& 0 \& 0 \& \cdots \& 0 \& 0 \& \ddots \& 0 \\
        0 \& 0 \& \ddots \& \ddots \& \ddots \& \ddots \& 0 \& \cdots \& 0 \&  0 \& \cdots \&  \alpha \\
        \vdots \& \vdots \&  \& 1 \&\ddots \&\ddots \& \ddots \& \& \vdots \& X \& \cdots \& X \\
        \vdots \& \vdots \&  \& \& \ddots \& X \& \cdots \& X \& 0 \& \vdots \& \& \vdots \\
        0 \& 0 \& \cdots \& \cdots \& 0 \& 1 \&  X \& \cdots \& X \& X \& \cdots \& X \\
        };
    \node[draw,color=blue,fit=(M-1-8)(M-6-10),inner sep=2pt,label={[xshift=0cm, yshift=0cm]\textcolor{blue}{$\svv{G}_{(0)}$}}] {};
    \node[draw,color=red,fit=(M-2-9)(M-6-11),inner sep=1pt,label={[xshift=0cm, yshift=-0.1cm]\textcolor{red}{$\svv{G}_{(1)}$}}] {};
    \node[draw,color=green,fit=(M-3-10)(M-6-12),inner sep=0pt,label={[xshift=0cm, yshift=0cm]}] {};
    \draw[decorate,decoration={brace,raise=7mm,amplitude=10pt}](M-1-12.north) -- node [right of=M,right=3mm] {$B-N+1$} (M-3-12.south);
    \draw[decorate,decoration={brace,raise=7mm,amplitude=10pt}](M-4-12.north) -- node [right of=M,right=3mm] {$T-B$} (M-6-12.south);
    \draw[decorate,decoration={brace,raise=4mm,amplitude=10pt,mirror}](M-6-1.west) -- node [below of=M,below=-3mm] {$k$} (M-6-6.east);
    \draw[decorate,decoration={brace,raise=4mm,amplitude=10pt,mirror}](M-6-7.west) -- node [below of=M,below=-3mm] {$N-1$} (M-6-9.east);
    \draw[decorate,decoration={brace,raise=4mm,amplitude=10pt,mirror}](M-6-10.west) -- node [below of=M,below=-3mm] {$B-N+1$} (M-6-12.east);
    }
    \end{aligned}.
    \label{eq:Goption2}
\end{equation}

We note that
\begin{align}
    \svv{\hat{H}}_{(l),2:k-l}= \left(\svv{M}^{-1}_{T-B,T-B}\svv{G}_{(l),T-B,:} \right)^T
\end{align}
where $\svv{M}^{-1}_{T-B,T-B}$ is acquired by taking the lower right $(T-B\times T-B)$ matrix of $\svv{M}^{-1}$, and $\svv{G}_{(l),T-B,:}$ is acquired by taking the last $T-B$ rows of $\svv{G}_{(l)}$. Since $\svv{M}^{-1}$ is an upper triangular matrix we have
\begin{align}
    {\rm rank}(\svv{M}^{-1}_{T-B,T-B})=T-B.
    \label{eq:rankM_min1_second}
\end{align}
Further, since taking the last $T-B$ rows of $\svv{G}_{(l)}$ results with a matrix which a sub-matrix of $\svv{G}_{T-B,B}$ (by taking only $T-B$ columns out of the $B$ columns of $\svv{G}_{T-B,B}$), recalling Property~\ref{prop:prop4}, it follows that
\begin{align}
    {\rm rank}(\svv{G}_{(l),T-B,:})=T-B.
    \label{eq:rankG_second}
\end{align}

Therefore, following Lemma~\ref{lem:multUppBoundByMin}, \eqref{eq:rankM_min1_second} and \eqref{eq:rankG_second},  we have that
\begin{align}
{\rm rank}(\svv{\hat{H}}_{(l),2:k-l})&\leq \min\left({\rm rank}(\svv{M}^{-1}_{T-B,T-B}),{\rm rank}(\svv{G}_{(l),T-B,:})\right) \nonumber \\
& = T-B.
\end{align}

Therefore, since ${\rm rank}(\svv{\hat{H}}_{(l)})=T-B+1$ and ${\rm rank}(\svv{\hat{H}}_{(l),2:k-l})\leq T-B$, it follows that ${\bf {\hat h}}_{(l),0}$ is linearly independent with the following $k-l+1$ columns.
\begin{enumerate}
\addtocounter{enumi}{1}
\item {\bf Condition R1}
\end{enumerate}
    We note that for $0\leq l \leq B-N$, we denote by $\svv{\bar{H}^{(l)}}$ as taking the last $T-l$ columns of $\svv{H}^{(l)}$. Hence showing that the $l$-th column, ${\bf h}_l^{(l)}$ of $\svv{H}^{(l)}$ is linearly independent of the any set of $(N-1)$ columns taken from the set
    \begin{align*}
        \left\{{\bf h}_j^{(l)},~l+1\leq j \leq l+T\right\}
    \end{align*}
    is equivalent to show that ${\bf{\bar{h}}}_0$ is linearly independent of any set of $(N-1)$ columns taken from the other $T-l-1$ columns of $\svv{\bar{H}^{(l)}}$. $\svv{\bar{H}^{(l)}}$ can be denoted as
    \begin{equation}
        \begin{aligned}
        \svv{\bar{H}^{(l)}}&=\svv{{H}}^{(l)}_{1:N+l,l:l+T} \nonumber \\
        &=
            \tikz[baseline=(M.west)]{%
        \node[matrix of math nodes,matrix anchor=west,left delimiter={[},right delimiter={]},ampersand replacement=\& ] (M) {
     -Y \& \cdots \& \cdots \&  \cdots \&  \cdots \&  \cdots \&  -Y \& 1 \& 0 \& \cdots \& \cdots \& \cdots \& 0 \\
    \vdots \& \vdots \& \vdots \&  \vdots \&  \vdots \&  \vdots \&  \vdots \& 0 \& \ddots \& \cdots \& \cdots \& \cdots \& 0 \\
    -Y \& \cdots \& \cdots \&  \cdots \&  \cdots \&  \cdots \&  -Y \& 0 \& \ldots \& 1 \& \vdots \& \cdots \& 0 \\
    -Y_{\alpha} \& \cdots \& \cdots \&  -Y_{\alpha} \&  -Y \&  \cdots \&  -Y \& \vdots \& \cdots \& \cdots \& 1 \& \vdots \& 0 \\
    \vdots \& \vdots \& \vdots \&  \vdots \&  \vdots \&  \vdots \&  \vdots \& \vdots \& \cdots \& \vdots \& \vdots \& \ddots \& 0 \\
    f_{l+1,l+1}(\alpha) \& -Y_{\alpha} \& \cdots \& -Y_{\alpha} \& -Y \& \cdots \& -Y \& \cdots \& \cdots \& \cdots \& \cdots \& \cdots \& 1 \\
            };
        \node[draw,color=blue,fit=(M-6-1)(M-1-13),inner sep=2pt,label=below:\textcolor{blue}{}] {};
        \node[draw,color=blue,densely dashed,fit=(M-1-2)(M-3-7),inner sep=0pt,label={[xshift=-1cm, yshift=0cm]}] {};
        \node[draw,color=blue,densely dashed,fit=(M-1-8)(M-3-10),inner sep=0pt,label={[xshift=-1cm, yshift=0cm]}] {};
        \node[draw,color=red,dash dot,fit=(M-1-2)(M-6-10),inner sep=2pt,label={[xshift=-1cm, yshift=0cm]}] {};
        \draw[decorate,decoration={brace,raise=4mm,amplitude=10pt,mirror}](M-6-1.west) -- node [below of=M,below=-3mm] {$B-N+1-l$} (M-6-4.east);
        \draw[decorate,decoration={brace,raise=4mm,amplitude=10pt,mirror}](M-6-5.west) -- node [below of=M,below=-3mm] {$T-B$} (M-6-7.east);
        \draw[decorate,decoration={brace,raise=4mm,amplitude=10pt,mirror}](M-6-8.west) -- node [below of=M,below=-3mm] {$N-1$} (M-6-10.east);
        \draw[decorate,decoration={brace,raise=4mm,amplitude=10pt,mirror}](M-6-11.west) -- node [below of=M,below=-3mm] {$l+1$} (M-6-13.east);
        \draw[decorate,decoration={brace,raise=11mm,amplitude=10pt,mirror}](M-6-1.west) -- node [below of=M,below=5mm] {$T-l$} (M-6-10.east);
        \draw[decorate,decoration={brace,raise=13mm,amplitude=10pt,mirror}](M-1-1.north) -- node [left of=M,left=6mm] {$N-1$} (M-3-1.south);
        \draw[decorate,decoration={brace,raise=25mm,amplitude=10pt,mirror}](M-1-1.north) -- node [left of=M,left=18mm] {$N+l$} (M-6-1.south);
        }
        \end{aligned}.
        \label{eq:HlCau_3}
    \end{equation}

Denote $\mathcal{A}\subset\{2,\ldots,T\}$ the set of $N-1$ coordinates taken from  $\svv{\bar{H}}^{(l)}_{:,2:T}$. Hence, we denote as $\svv{\bar{H}}^{(l)}_{:,\mathcal{A}}$ as the sub matrix formed from taking the columns matching these coordinates (i.e. the matrix of $N-1$ columns we need to show that ${\bf \bar{h}}_0$ is linearly independent of). 

We note that $\svv{\bar{H}}^{(l)}$ is a $N+l\times T+1$ matrix therefore $\svv{\bar{H}}^{(l)}_{:,\mathcal{A}}$ is a $N+l\times N-1$ matrix. Hence
\begin{align}
    {\rm rank}(\svv{\bar{H}}^{(l)}_{:,\mathcal{A}})\leq N-1.
\end{align}

Next, we show that
\begin{align}
    {\rm rank}(\svv{\bar{H}}^{(l)}_{:,\{1,\mathcal{A}\}})=N,
\end{align}
i.e. that the rank of all $N$ columns equals $N$.

We note that the $N-1\times T-l-1$ upper right sub matrix of $\svv{\bar{H}}^{(l)}$, can be denoted as
\begin{align}
    \svv{\bar{H}}^{(l)}_{1:N-1,2:,T-l}=\begin{bmatrix}
    \svv{\bar{P}} & I_{N-1} \end{bmatrix},
\end{align}
where $\svv{\bar{P}}$ is a $N-1\times k-1-l$ sub matrix of $\svv{P}''$. This is the dash-dot matrix in \eqref{eq:HlCau_3}.

Denote the set of $w_1$ vectors taken from indices $\{2,\ldots,k-l\}$ as  $\mathcal{W}_1\subset\{2,\ldots,k-l\}$, the set of $w_2$ vectors taken from indices $\{k-l+1,\ldots,T-l\}$ as $\mathcal{W}_2\subset\{k-l+1,\ldots,T-l\}$. We note that $\mathcal{W}=\mathcal{W}_1\cup \mathcal{W}_1$ is the set of $w=w_1+w_2$ vectors taken from indices $\{2,\ldots,T-l\}$. We further denote the set  $\mathcal{\tilde{W}}_2\subset\{2,\ldots,N-1\}$ where this set indicates the location of the elements in $\mathcal{W}_2$ with respect to column $K-l+1$.

Taking columns in $\mathcal{W}$ from $\svv{\bar{H}}^{(l)}_{1:N-1,2:,T-l-1}$ can be denoted as
\begin{align}
    \svv{\bar{H}}^{(l)}_{1:N-1,\mathcal{W}}=\begin{bmatrix}
    \svv{\bar{P}}_{\mathcal{W}_1} & I_{N-1,\mathcal{W}_2} \end{bmatrix}.
\end{align}

We show next that 
\begin{align}
    {\rm rank}(\svv{\bar{H}}^{(l)}_{1:N-1,\mathcal{W}})=w.
\end{align}

Applying column operations, we can zero the $w_2$ rows corresponds to indices $\mathcal{\tilde{W}}_2$ in $\svv{\bar{P}}_{\mathcal{W}_1}$. Thus, we are left with a matrix composed taking columns in indices $\mathcal{W}_1$ and rows in $\{1:N-1\}\cap \mathcal{\tilde{W}}_2$. Following Corollary~\ref{col:rankOFpartialP}, this matrix has rank which equals $\min\left(N-1-w_2,w_1\right)$.

An example where $w_1=3$ and $w_2=2$ is given in \eqref{eq:ExampleOfW} (after applying some row swaps) .
    \begin{equation}
        \begin{aligned}
        \svv{\bar{H}}^{(l)}_{1:N-1,\mathcal{W}} &=\tikz[baseline=(M.west)]{%
        \node[matrix of math nodes,matrix anchor=west,left delimiter={[},right delimiter={]},ampersand replacement=\& ] (M) {
     -Y \&  -Y \&   -Y \& 1 \& 0  \\
    \vdots \& \vdots \& \vdots \& 0 \& 1  \\
    \vdots \& \vdots  \& \vdots \& 0 \& 0 \\
    \vdots \& \vdots \&  \vdots \& \vdots \& \vdots  \\
    \vdots \& \vdots \&  \vdots \& \vdots \& \vdots  \\
    \vdots \& \vdots \&  \vdots \& \vdots \& \vdots  \\
    \vdots \& \vdots \&  \vdots \& \vdots \& \vdots  \\
    -Y \&  \cdots \&  -Y \& 0 \& 0  \\
    };
        } =\tikz[baseline=(M.west)]{%
        \node[matrix of math nodes,matrix anchor=west,left delimiter={[},right delimiter={]},ampersand replacement=\& ] (M) {
     0 \&  0 \&   0 \& 1 \& 0  \\
    0 \& 0 \& 0 \& 0 \& 1  \\
    -Y \& -Y  \& -Y \& 0 \& 0  \\
    \vdots \& \vdots \&  \vdots \& \vdots \& \vdots  \\
    \vdots \& \vdots \&  \vdots \& \vdots \& \vdots  \\
    \vdots \& \vdots \&  \vdots \& \vdots \& \vdots  \\
    \vdots \& \vdots \&  \vdots \& \vdots \& \vdots  \\
    -Y \&  \cdots \&  -Y \& 0 \& 0  \\
    };
        } =\tikz[baseline=(M.west)]{%
        \node[matrix of math nodes,matrix anchor=west,left delimiter={[},right delimiter={]},ampersand replacement=\& ] (M) {
     0 \&  0 \&   0 \& 1 \& 0  \\
    0 \& 0 \& 0 \& 0 \& 1  \\
    1 \& 0  \& 0 \& 0 \& 0  \\
    0 \& 1 \& 0 \& \vdots \& \vdots  \\
    0 \& 0 \& 1 \& \vdots \& \vdots  \\
    -\tilde{Y} \&  \cdots \&  -\tilde{Y} \& 0 \& 0  \\
    \vdots \& \vdots \&  \vdots \& \vdots \& \vdots  \\
    -\tilde{Y} \&  \cdots \&  -\tilde{Y} \& 0 \& 0  \\
    };
        }.
        \end{aligned}
        \label{eq:ExampleOfW}
    \end{equation}
    
Therefore, applying column operations (and some row swaps), we get that 
\begin{align}
    \svv{\bar{H}}^{(l)}_{:,\mathcal{W}}=\begin{bmatrix}
    I_{w} \\ 
    \svv{\tilde{H}}^{(l)}_{w+1:N+l,\mathcal{W}} \end{bmatrix},
\end{align}
where $\svv{\tilde{H}}^{(l)}_{w+1:N-1,\mathcal{W}}$ are the $N+l-w$ rows resulting from the column operations (over $\mathbb{F}_q$).

We note that if $w=N-1$ (i.e., none of the erasures occurred in indices ${\{T-l+1,\ldots,T+1\}}$) we have $\mathcal{A}=\mathcal{W}$, thus we have 
\begin{align}
    \svv{\bar{H}}^{(l)}_{:,\{1,\mathcal{A}\}}= \begin{bmatrix}
    {\bf 0}_{N-1\times 1} & I_{N-1} \\ 
    {\bf{\tilde h}}_{0,l+1\times 1} & \svv{\tilde{H}}^{(l)}_{w+1:N+l,\mathcal{W}} \end{bmatrix}
\end{align}
where ${\bf{\tilde h}}_{0,l+1\times 1}$ is the $l+1\times 1$ lower part of ${\bf{\tilde h}_0}$ which is the outcome of the column operations that zeroed the upper $N-1$  part of this vector. Following Property~\ref{prop:f_ii_not_zero} we have that $f_{l+1,l+1}(\alpha)\neq 0$ and since all column operations were performed over $\mathbb{F}_q$ it follows that $\tilde{f}_{l+1,l+1}(\alpha)\neq 0$. Thus, we have that 
\begin{align}
    {\rm rank}(\svv{\bar{H}}^{(l)}_{:,\{1,\mathcal{A}\}})=N.
\end{align}

We note that if $w<N-1$ (i.e. some of the erasure occurred at the last $l+1$ indices), when applying column operations, not all upper $N-1$ elements of ${\bf{\tilde h}}_{0,l+1\times 1}$ are nulled. Since these elements are taken from $\svv{P}''$, it follows each one of them does not equal 0. Thus, again we get that 
\begin{align}
    {\rm rank}(\svv{\bar{H}}^{(l)}_{:,\{1,\mathcal{A}\}})=N.
\end{align}

\begin{enumerate}
\addtocounter{enumi}{2}
\item {\bf Condition B2}
\end{enumerate}
We note that for $B-N+1\leq l \leq T-N+1$, the set
    \begin{align*}
        \left\{{\bf h}_j,~l\leq j \leq l+B-1\right\}
    \end{align*}
can be viewed as set of $B$ columns taken from the parity-check matrix associated with the $(n,k)$ MDS code $\mathcal{C}''$ with generator matrix $\svv{G}''$. Examples of groups of $B$ columns to be analyzed are depicted in \eqref{eq:Hb2} below for $l=B-N+1$ and $l=B-N+2$.

\begin{equation}
\renewcommand\arraystretch{2}
\begin{aligned}
\svv{H}&=\left[-\svv{\tilde{P}}^T\given[\Big]\svv{I}_{B\times B}\right] \\
&=\tikz[baseline=(M.west)]{%
    \node[matrix of math nodes,matrix anchor=west,left delimiter={[},right delimiter={]},ampersand replacement=\& ] (M) {
    -Y \& \cdots \& \cdots \&  \cdots \&  \cdots \&  \cdots \& -Y \& 1 \& 0 \& \cdots \& \cdots \& \cdots \& \cdots \& 0   \\
\vdots \& \vdots \& \vdots \&  \vdots \&  \vdots \&  \vdots \& \vdots \& 0 \& 1 \& 0 \& \cdots \& \cdots \& \cdots \& 0 \\
-Y \& \cdots \& \cdots \&  \cdots \&  \cdots \&  \cdots \&  -Y \& 0 \& 0 \& \ddots \& \cdots \& \vdots \& \vdots \& 0    \\
f_{1,1}(\alpha) \& -Y_{\alpha} \& \cdots \& -Y_{\alpha} \& \cdots \& \cdots \& -Y \& \vdots \& \vdots \&  \& 1 \& \ddots \&  \vdots \& 0    \\
f_{1,2}(\alpha) \& f_{2,2}(\alpha) \& -Y_{\alpha} \& \vdots \& \cdots \& \cdots \& -Y \& \vdots \& \vdots \&  \& 0 \& \ddots \& \vdots \& 0    \\
\vdots \& \vdots \& \ddots \& -Y_{\alpha} \&\vdots \&\vdots  \& -Y \& \vdots \& \vdots \&  \& \& 0 \&  \ddots \& 0    \\
f_{1,B-N+1}(\alpha) \& f_{2,B-N+1}(\alpha) \& \cdots \& f_{B-N+1,B-N+1}(\alpha) \&-Y \&\cdots  \& -Y \& 0 \& 0 \& \cdots \& \cdots \& 0 \& 0 \& 1 \\
};
    \node[draw,color=blue,fit=(M-1-5)(M-7-14),inner sep=1pt,label=below:\textcolor{blue}{}] {};
    \draw[decorate,decoration={brace,raise=4mm,amplitude=10pt,mirror}](M-7-5.west) -- node [below of=M,below=-3mm] {$N-1$} (M-7-7.east);
    }
    \end{aligned}
    \label{eq:Hb2}
\end{equation}

Following Property~\ref{def:Cauchy}, we note that $\svv{H}_{:,\{B-N+2,\ldots,n\}}$ (marked in solid in \eqref{eq:Hb2}), can be viewed as the parity matrix of $(B+N-1,N-1)$ MDS code. Following Corollary~\ref{col:anyNminKIsInd}, any $B$ columns from the parity of this code matrix are independent, it follows that for any $B-N+1\leq l \leq T-N+1$, the set $\left\{{\bf h}_j,~l\leq j \leq l+B-1\right\}$  is independent.



\begin{enumerate}
\addtocounter{enumi}{3}
\item {\bf Condition R2}
\end{enumerate}
Following the explanation for condition B2, and since $B\geq N$, it follows that condition R2 also holds.

\section{Concluding remarks}
In this paper, we study streaming codes over an erasure channel
whose erasure pattern in every sliding window of size $W$ is either
a burst erasure of maximum length $B$ or multiple arbitrary erasures
of maximum total count $N$ where these parameters meet
\eqref{eq:WTBN}. While the capacity of this channel was derived
(separately) by Fong et al. in \cite{fong2019optimal} and Krishnan
et al. \cite{krishnan2018rate}, no explicit code construction which
achieves this capacity with low field size (for any coding
parameters) was suggested.

In this paper, we presented an explicit code construction that achieves the capacity of the channel mentioned above with a field size that scales quadratically in the delay constraint. The construction relays on properties of MDS codes from a base field (whose size scales linearly with the delay constraint) while replacing some of the elements in the generator matrix with elements for an extension field, thus resulting in a final field size that scales quadratically with the delay constraint.

An open question is whether there exists an explicit construction for capacity-achieving streaming codes (for all admissible parameters) with smaller field size (i.e., a field size the scales linearly with the delay constraint). While this seems possible to some specific values of $(W,B,N)$, finding a coding scheme for any (admissible) values is an interesting avenue.

\appendices
\section{Explicit description of $\svv{\tilde{G}}$}
\label{app:explicitDescOfTildeG}
Writing \eqref{eq:tildeG} explicitly we have
\begin{equation}
\begin{aligned}
    \svv{\tilde{G}}& =\svv{M}^{-1}\svv{G} \nonumber \\
    & = \begin{bmatrix} \svv{M}^{-1}_{1,:} \\ \svv{M}^{-1}_{2,:} \\ \vdots \\ \vdots \\ \vdots \\ \svv{M}^{-1}_{k,:} \end{bmatrix} \begin{bmatrix} \svv{G}_{:,1} \cdots \svv{G}_{:,n} \end{bmatrix}
    \nonumber \\
    &= \begin{bmatrix} \svv{M}^{-1}_{1,:} \\ \svv{M}^{-1}_{2,:} \\ \vdots \\ \vdots \\ \vdots \\ \svv{M}^{-1}_{k,:} \end{bmatrix}
    \tikz[baseline=(M.west)]{%
    \node[matrix of math nodes,matrix anchor=west,left delimiter={[},right delimiter={]},ampersand replacement=\& ] (M) {
        1 \& X \& \cdots \& X \& 0 \& 0 \& 0 \& \cdots \& 0 \& \alpha \& \cdots \& 0 \\
        0 \& 1 \& X \& \cdots \& X \& 0 \& 0 \& \cdots \& 0 \& 0 \& \ddots \& 0 \\
        0 \& 0 \& \ddots \& \ddots \& \ddots \& \ddots \& 0 \& \cdots \& 0 \&  0 \& \cdots \&  \alpha \\
        \vdots \& \vdots \&  \& 1 \&\ddots \&\ddots \& \ddots \& \& \vdots \& X \& \cdots \& X \\
        \vdots \& \vdots \&  \& \& \ddots \& X \& \cdots \& X \& 0 \& \vdots \& \& \vdots \\
        0 \& 0 \& \cdots \& \cdots \& 0 \& 1 \&  X \& \cdots \& X \& X \& \cdots \& X \\
        };
    \draw[decorate,decoration={brace,raise=7mm,amplitude=10pt}](M-1-12.north) -- node [right of=M,right=3mm] {$B-N+1$} (M-3-12.south);
    \draw[decorate,decoration={brace,raise=4mm,amplitude=10pt}](M-1-10.west) -- node [above of=M,above=-3mm] {$B-N+1$} (M-1-12.east);
    } \nonumber \\
    & = \tikz[baseline=(M.west)]{%
    \node[matrix of math nodes,matrix anchor=west,left delimiter={[},right delimiter={]},ampersand replacement=\& ] (M) {
    1 \& 0 \& \cdots \& \cdots \& \cdots \& \cdots \& 0 \&  Y \& \cdots \& Y \& f_{1,1}(\alpha) \& \cdots \& \cdots \& f_{1,B-N+1}(\alpha) \\
    0 \& 1 \& 0 \& \cdots \& \cdots \& \cdots \& 0 \& Y \& \cdots \& Y \&  Y_{\alpha} \& f_{2,2}(\alpha) \&  \cdots \& f_{2,B-N+1}(\alpha) \\
    0 \& 0 \& \ddots \& \cdots \& \vdots \& \vdots \& 0 \& Y \& \cdots \& Y \& \vdots \& \ddots \& \ddots \& \vdots     \\
    \vdots \& \vdots \&  \& \ddots \&\vdots \& \vdots \& 0 \& Y \& \cdots \& Y  \&  Y_{\alpha} \& \cdots \&  Y_{\alpha} \& f_{B-N+1,B-N+1}(\alpha)      \\
    \vdots \& \vdots \&  \& \& \ddots \&  \vdots \& 0 \& Y \& \cdots \& Y  \&  Y \& \cdots \& Y \& Y \\
    0 \& 0 \& \cdots \& \cdots \& 0 \& 1 \& 0 \&  Y \& \cdots \& Y \& Y \& \cdots \& Y \& Y \\
    0 \& 0 \& \cdots\& \cdots \&  \cdots  \& 0 \& 1 \& Y \& \cdots \& Y  \& Y \& \cdots \& Y  \& Y \\
    };
    \draw[decorate,decoration={brace,raise=4mm,amplitude=10pt,mirror}](M-7-1.west) -- node [below of=M,below=-3mm] {$k$} (M-7-7.east);
    \draw[decorate,decoration={brace,raise=4mm,amplitude=10pt,mirror}](M-7-8.west) -- node [below of=M,below=-3mm] {$N-1$} (M-7-10.east);
    \draw[decorate,decoration={brace,raise=4mm,amplitude=10pt,mirror}](M-7-11.west) -- node [below of=M,below=-3mm] {$B+N-1$} (M-7-14.east);
    }.
\end{aligned}
\end{equation}

Recalling that $\svv{M}$ is an upper triangular matrix it follows that $\svv{M}^{-1}$ is also upper triangular. Since $\svv{G}_{:,1:T} = \svv{G}'_{:,1:T}$ it follows that the first $T$ columns of $\svv{\tilde{G}}$ equal the first $T$ columns of $\svv{G}'$. 

We denote with $f_{i,j}=\svv{M}^{-1}_{i,:}\svv{G}_{j-T,:}$ where $1\leq i \leq B+N-1$ and $T+i \leq j \leq n$. We note that only these elements in $\svv{G}$ include $\alpha\in\mathbb{F}_{q^2}\setminus\mathbb{F}_q$. We further note that since $\svv{M}^{-1}_{i,i}=1$ (for any $i\in\{1,\ldots,k\}$, this element is not nulled in this multiplication. 

We denote with $Y_{\alpha}=\svv{M}^{-1}_{i,:}\svv{G}_{j-T,:}$ where  $1\leq i \leq B+N-1$ and $T+1 \leq j \leq T+i$, i.e., these elements does not equal to the elements at this location in $\svv{G}'$ (due to replacing the upper right $B-N+1\times B-N+1$ matrix in $\svv{G}'$ with $\svv{I}\cdot{\alpha}$). We note though that these elements does not contain $\alpha$.
\section{Proof of Lemma~\ref{lem:multUppBoundByMin}}
\label{app:proofPfLemUpp}
The space spanned by the columns of $\svv{A}\svv{B}$ is the space $\mathcal{S}$ of all vectors $s$ that can be written as linear combinations of the columns of $\svv{A}\svv{B}$:
\begin{align}
    s=\left(\svv{A}\svv{B}\right)v
\end{align}
where $v$ is the $M\times 1$ vector of coefficients of the linear combination. We can also write
\begin{align}
    s=\svv{A}\left(\svv{B}v\right)
\end{align}
where $\svv{B}v$ is an $L\times 1$ vector (being a product of an $L\times M$ matrix and an $M \times 1$ vector). Thus, any vector $s\in \mathcal{S}$ can be written as a linear combination of the columns of $\svv{A}$, with coefficients taken from the vector $\svv{B}v$. As a consequence, the space $\mathcal{S}$ is no larger than the span of the columns of $\svv{A}$, whose dimension is ${\rm rank}(\svv{A})$. This implies that the dimension of $\mathcal{S}$ is less than or equal to ${\rm rank}(\svv{A})$. Since the dimension of $\mathcal{S}$ is the rank of $\svv{A}\svv{B}$, we have
\begin{align}
    {\rm rank}(\svv{A}\svv{B})\leq{\rm rank}(\svv{A})
    \label{eq:rankABlowerRankA}
\end{align}
Now, the space spanned by the rows of $AB$ is the space $\mathcal{T}$ of all vectors $t$ that can be written as linear combinations of the rows of $\svv{A}\svv{B}$:
\begin{align}
    t=u\left(\svv{A}\svv{B}\right)
\end{align}
where $u$ is the $1\times K$ vector of coefficients of the linear combination. We can also write
\begin{align}
    t=\left(u\svv{A}\right)\svv{B}
\end{align}
where $u\svv{A}$ is a $1\times L$ vector (being a product of a $1\times K$ vector and a $K\times L$ matrix). Thus, any vector $t\in\mathcal{T}$ can be written as a linear combination of the rows of $\svv{B}$, with coefficients taken from the vector $u\svv{A}$. As a consequence, the space $\mathcal{T}$ is no larger than the span of the rows of $\svv{B}$, whose dimension is ${\rm rank}(\svv{B})$. This implies that the dimension of $\mathcal{T}$ is less than or equal to ${\rm rank}(\svv{B})$. Since the dimension of $T$ is the rank of $\svv{A}\svv{B}$, we have
\begin{align}
    {\rm rank}(\svv{A}\svv{B})\leq{\rm rank}(\svv{B})
    \label{eq:rankABlowerRankB}
\end{align}

\eqref{eq:rankABlowerRankA} and \eqref{eq:rankABlowerRankB} imply that
\begin{align}
    {\rm rank}(\svv{A}\svv{B})\leq \min\left( {\rm rank}(\svv{A}),{\rm rank}(\svv{B})\right)
\end{align}
\section{Proof of Proposition~\ref{prop:prop4}}
\label{sec:app_prop4}
\begin{proof}
Recall that $\svv{G}_{T-B,B}$ the lower right $(T-B\times B)$ matrix of $\svv{G}$ marked in solid in \eqref{eq:G_TB}.

\begin{equation}
    \begin{aligned}
    \svv{G}=
        \tikz[baseline=(M.west)]{%
    \node[matrix of math nodes,matrix anchor=west,left delimiter={[},right delimiter={]},ampersand replacement=\& ] (M) {
        1 \& X \& \cdots \& X \& 0 \& 0 \& 0 \& \cdots \& 0 \& \alpha \& \cdots \& 0 \\
        0 \& 1 \& X \& \cdots \& X \& 0 \& 0 \& \cdots \& 0 \& 0 \& \ddots \& 0 \\
        0 \& 0 \& \ddots \& \ddots \& \ddots \& \ddots \& 0 \& \cdots \& 0 \&  0 \& \cdots \&  \alpha \\
        \vdots \& \vdots \&  \& 1 \&\ddots \&\ddots \& \ddots \& \& \vdots \& X \& \cdots \& X \\
        \vdots \& \vdots \&  \& \& \ddots \& X \& \cdots \& X \& 0 \& \vdots \& \& \vdots \\
        0 \& 0 \& \cdots \& \cdots \& 0 \& 1 \&  X \& \cdots \& X \& X \& \cdots \& X \\
        };
    \node[draw,color=blue,fit=(M-6-7)(M-4-12),inner sep=0pt,label=below:{$\svv{G}_{T-B,B}$}] {};
    \draw[decorate,decoration={brace,raise=7mm,amplitude=10pt}](M-4-12.north) -- node [right of=M,right=3mm] {$T-B$} (M-6-12.south);
    }
    \end{aligned}.
    \label{eq:G_TB}
\end{equation}

First step in analyzing the rank of $\svv{G}_{T-B,B}$ is to note that $\svv{G}_{T-B,B}=\svv{G}'_{T-B,B}$. Now, from the definition of $\svv{G}'$ we have that $\svv{G}'=\svv{M}\svv{G}''$. Hence, $\svv{G}_{T-B,B}$ can be viewed as multiplying the $(T-B\times T-B)$ lower right part of $\svv{M}$ (which we denote as $\svv{M}_{T-B,T-B}$) with the $(T-B \times B)$ lower right part of $\svv{G}''$ (which we denote as $\svv{G}''_{T-B,B}$).
\begin{align}
    \svv{G}_{T-B,B}&=\svv{M}_{T-B,T-B}\svv{G}''_{T-B,B} \nonumber \\
    & = \underbrace{\tikz[baseline=(M.west)]{%
    \node[matrix of math nodes,matrix anchor=west,left delimiter={[},right delimiter={]},ampersand replacement=\& ] (M) {
    1 \& Y' \& \cdots \& Y' \& 0 \& \cdots \& \cdots \& 0 \\
    0 \& 1 \& Y' \& \cdots \& Y' \& 0 \& \cdots \& 0 \\
    \vdots \& \cdots \& \ddots \& \ddots \& \ddots \& \ddots \& \ddots \& \vdots\\
    0 \& \cdots \& 0 \& 1 \& Y' \& \cdots \& Y' \& 0 \\
    0 \& \cdots \& 0 \& 0 \& 1 \& Y' \& \cdots \& Y' \\
    0 \& \cdots \& 0 \& 0 \& 0 \& \ddots \& \ddots \& Y' \\
    0 \& \cdots \& 0 \& 0 \& 0 \& 0 \& 1 \& Y' \\
    0 \& \cdots \& 0 \& 0 \& 0 \& 0 \& 0 \& 1 \\
};
\node[draw,color=blue,fit=(M-5-5)(M-8-8),inner sep=1pt,label={[xshift=-1cm, yshift=0cm]}] {};
}}_{\svv{M}}
    \underbrace{\tikz[baseline=(M.west)]{%
    \node[matrix of math nodes,matrix anchor=west,left delimiter={[},right delimiter={]},ampersand replacement=\& ] (M) {
    1 \& 0 \& \cdots \& 0 \& 0 \& \cdots \& \cdots \& 0 \& Y \& \cdots \& Y \\
    0 \& 1 \& 0 \& \cdots \& 0 \& 0 \& \cdots \& 0 \& Y \& \cdots \& Y \\
    \vdots \& \cdots \& \ddots \& \vdots \& \vdots \& \vdots \& \vdots \& \vdots \& \vdots \& \vdots \& \vdots \\
    0 \& \cdots \& 0 \& 1 \& 0 \& \cdots \& 0 \& 0 \& Y \& \cdots \& Y \\
    0 \& \cdots \& 0 \& 0 \& 1 \& 0 \& \cdots \& 0 \& Y \& \cdots \& Y \\
    0 \& \cdots \& 0 \& 0 \& 0 \& \ddots \& \ddots \& 0 \& Y \& \cdots \& Y \\
    0 \& \cdots \& 0 \& 0 \& 0 \& 0 \& 1 \& 0 \& Y \& \cdots \& Y \\
    0 \& \cdots \& 0 \& 0 \& 0 \& 0 \& 0 \& 1 \& Y \& \cdots \& Y\\
};
\node[draw,color=blue,dashed,fit=(M-1-9)(M-8-11),inner sep=2pt,label={$\svv{P}''_{k\times B}$}] {};
\node[draw,color=blue,fit=(M-5-9)(M-8-11),inner sep=0pt,label={[xshift=-1cm, yshift=0cm]}] {};
}}_{\svv{G}''}
\end{align}

Since $\svv{G}''_{T-B,B}$ is a sub-matrix of $\svv{P}''$, following Corollary~\ref{col:CauchyForRectangular}, it has a full rank. Thus,
\begin{align}
    {\rm rank}(\svv{G}''_{T-B,B})= \min(T-B,B).
\end{align}

Matrix $\svv{M_{T-B,T-B}}$ is a square upper triangular matrix and thus we have
\begin{align}
     {\rm rank}(\svv{M}_{T-B,T-B})=T-B.
\end{align}

Following Lemma~\ref{lem:multUppBoundByMin} we have that
\begin{align}
    {\rm rank}(\svv{G}_{T-B,B})\leq & \min\left({\rm rank}(\svv{M}_{T-B,T-B}),{\rm rank}(\svv{G}''_{T-B,B})\right) \nonumber \\
    & \min(T-B,B).
    \label{eq:GrankUpp}
\end{align}

Further, following Corollary~\ref{col:colSylFullRank} (and recalling that $\svv{M}_{T-B,T-B}$ is a square, full rank matrix) we have
\begin{align}
    {\rm rank}(\svv{G}_{T-B,B})& \geq {\rm rank}(\svv{G}''_{T-B,B}) \nonumber \\
    &= \min(T-B,B).
    \label{eq:GrankLow}
\end{align}

Combining \eqref{eq:GrankUpp} and \eqref{eq:GrankLow}, we conclude that ${\rm rank}(\svv{G}_{T-B,B})=\min(T-B,B)$.

Further, we note that the sub-matrix of $\svv{G}_{T-B,B}$ which is formed by taking any $w$ columns of $\svv{G}_{T-B,B}$ has a rank of $\min(T-B,w)$ since taking $w$ columns is equivalent to multiply $\svv{M}_{T-B,T-B}$ (which is a square full matrix) with $w$ columns from $\svv{G}''_{T-B,B}$.

\end{proof}
\section{Proof of Proposition~\ref{prop:prop5}}
\label{sec:app_prop5}
\begin{proof}
Recall that $\svv{M}^{-1}_{B,T-B}$ is the upper right $(B\times T-B)$ matrix of $\svv{M}^{-1}$ of marked in solid in \eqref{eq:invM_TmBCol}.

\begin{equation}
    \renewcommand\arraystretch{1.8}
    \svv{M}^{-1}=
    \tikz[baseline=(M.west)]{%
    \node[matrix of math nodes,matrix anchor=west,left delimiter={[},right delimiter={]},ampersand replacement=\& ] (M) {
    1 \& Y'' \& \cdots \& Y'' \& Y'' \& \cdots \& \cdots \& Y'' \\
    0 \& 1 \& Y'' \& \cdots \& Y'' \& Y'' \& \cdots \& Y'' \\
    \vdots \& \cdots \& \ddots \& \ddots \& \ddots \& \ddots \& \ddots \& \vdots\\
    0 \& \cdots \& 0 \& 1 \& Y'' \& \cdots \& Y'' \& Y'' \\
    0 \& \cdots \& 0 \& 0 \& 1 \& Y'' \& \cdots \& Y'' \\
    \vdots \& \cdots \& \vdots \& \vdots \& \vdots \& \ddots \& \ddots \& Y'' \\
    0 \& \cdots \& 0 \& 0 \& 0 \& 0 \& 1 \& Y'' \\
    0 \& \cdots \& 0 \& 0 \& 0 \& 0 \& 0 \& 1 \\
};
    \node[draw,color=blue,fit=(M-1-4)(M-5-8),inner sep=2pt,label=$\svv{M}^{-1}_{B,T-B}$] {};
    \draw[decorate,decoration={brace,raise=7mm,amplitude=10pt}](M-1-8.north) -- node [right of=M,right=3mm] {$B-N+1$} (M-3-8.south);
    \draw[decorate,decoration={brace,raise=7mm,amplitude=10pt}](M-4-8.north) -- node [right of=M,right=3mm] {$N-1$} (M-5-8.south);
    }
    \label{eq:invM_TmBCol}
\end{equation}

Since $\svv{M}\svv{M}^{-1}=\svv{I}$ we have
\begin{align}
    \underbrace{\tikz[baseline=(M.west)]{%
    \node[matrix of math nodes,matrix anchor=west,left delimiter={[},right delimiter={]},ampersand replacement=\& ] (M) {
    1 \& Y' \& \cdots \& Y' \& 0 \& \cdots \& \cdots \& 0 \\
    0 \& 1 \& Y' \& \cdots \& Y' \& 0 \& \cdots \& 0 \\
    \vdots \& \cdots \& \ddots \& \ddots \& \ddots \& \ddots \& \ddots \& \vdots\\
    0 \& \cdots \& 0 \& 1 \& Y' \& \cdots \& Y' \& 0 \\
    0 \& \cdots \& 0 \& 0 \& 1 \& Y' \& \cdots \& Y' \\
    0 \& \cdots \& 0 \& 0 \& 0 \& \ddots \& \ddots \& Y' \\
    0 \& \cdots \& 0 \& 0 \& 0 \& 0 \& 1 \& Y' \\
    0 \& \cdots \& 0 \& 0 \& 0 \& 0 \& 0 \& 1 \\
};
\node[draw,color=blue,fit=(M-1-1)(M-3-5),inner sep=1pt,label=$\svv{M}_{1:B-N+1,1:B}$]{};
}}_{\svv{M}}
    \underbrace{\tikz[baseline=(M.west)]{%
    \node[matrix of math nodes,matrix anchor=west,left delimiter={[},right delimiter={]},ampersand replacement=\& ] (M) {
    1 \& Y'' \& \cdots \& Y'' \& Y'' \& \cdots \& \cdots \& Y'' \\
    0 \& 1 \& Y'' \& \cdots \& Y'' \& Y'' \& \cdots \& Y'' \\
    \vdots \& \cdots \& \ddots \& \ddots \& \ddots \& \ddots \& \ddots \& \vdots\\
    0 \& \cdots \& 0 \& 1 \& Y'' \& \cdots \& Y'' \& Y'' \\
    0 \& \cdots \& 0 \& 0 \& 1 \& Y'' \& \cdots \& Y'' \\
    0 \& \cdots \& 0 \& 0 \& 0 \& \ddots \& \ddots \& Y'' \\
    0 \& \cdots \& 0 \& 0 \& 0 \& 0 \& 1 \& Y'' \\
    0 \& \cdots \& 0 \& 0 \& 0 \& 0 \& 0 \& 1 \\
};
\node[draw,color=blue,fit=(M-1-4)(M-5-8),inner sep=1pt,label=$\svv{M}^{-1}_{B,T-B}$]{};
}}_{\svv{M}^{-1}} =
 \tikz[baseline=(M.west)]{%
    \node[matrix of math nodes,matrix anchor=west,left delimiter={[},right delimiter={]},ampersand replacement=\& ] (M) {
    1 \& 0 \& \cdots \& 0 \& 0 \& \cdots \& \cdots \& 0 \\
    0 \& 1 \& 0 \& \cdots \& 0 \& 0 \& \cdots \& 0 \\
    \vdots \& \cdots \& \ddots \& \ddots \& \ddots \& \ddots \& \ddots \& \vdots\\
    0 \& \cdots \& 0 \& 1 \& 0 \& \cdots \& 0 \& 0 \\
    0 \& \cdots \& 0 \& 0 \& 1 \& 0 \& \cdots \& 0 \\
    0 \& \cdots \& 0 \& 0 \& 0 \& \ddots \& \ddots \& 0 \\
    0 \& \cdots \& 0 \& 0 \& 0 \& 0 \& 1 \& 0 \\
    0 \& \cdots \& 0 \& 0 \& 0 \& 0 \& 0 \& 1 \\
};
\node[draw,color=blue,fit=(M-1-4)(M-3-8),inner sep=1pt,label={[xshift=-1cm, yshift=0cm]}] {};
}
\end{align}

which means
\begin{align}
    \svv{M}_{1:B-N+1,1:B}\times \svv{M}^{-1}_{B,T-B} = \svv{0}_{B-N+1\times B}
\end{align}
which means that the first $B-N+1$ rows of $\svv{M}^{-1}_{B,T-B}$ can be expressed as linear combination of the last $N-1$ rows. 
\end{proof}

\bibliographystyle{IEEEtran}
\bibliography{eladd.bib}

\end{document}